%% file: main.tex
\documentclass[%
 aps, pra, reprint,
 superscriptaddress,
 longbibliography, nopreprintnumbers, noeprint,
 amsmath, amssymb, floatfix
]{revtex4-2}

\usepackage{multirow}
\usepackage{graphicx}
\usepackage{dcolumn}
\usepackage{bm}
\usepackage{dsfont}
\usepackage{hyperref}
\usepackage{physics}
\usepackage[usenames,dvipsnames]{xcolor}
\usepackage{bbold}
\usepackage[normalem]{ulem} 
\usepackage[inline]{enumitem}
\usepackage{qcircuit} 
\usepackage{tikz}
\usepackage{amssymb} 
\usepackage[caption=false]{subfig}
\usepackage{algorithm} 
\usepackage{algpseudocode} 
\usepackage{mathrsfs}
\usepackage{circuitikz} 
\DeclareMathAlphabet{\mathpzc}{OT1}{pzc}{m}{it}
\usepackage{booktabs} 

\newcommand{\sx}{\hat{\sigma}_x}
\newcommand{\sy}{\hat{\sigma}_y}
\newcommand{\sz}{\hat{\sigma}_z}

\newcommand{\h}{\hat{\mathcal{H}}}
\newcommand{\kHz}{~\mathrm{kHz}}

\newcommand{\figref}[1]{Fig.~\ref{#1}}
\newcommand{\secref}[1]{Section \ref{#1}}
\newcommand{\appref}[1]{Appendix \ref{#1}}
\renewcommand{\eqref}[1]{Eq.~\ref{#1}}

\newcommand{\lmin}{L_\mathrm{min}}
\newcommand{\lambdall}{\lambda_\mathrm{ll}}

\begin{document}

\title{Fast-tracking and disentangling of qubit noise fluctuations using minimal-data averaging and hierarchical discrete fluctuation auto-segmentation}

\author{Abhishek Agarwal}
\email{abhishek.agarwal@npl.co.uk}
\author{Lachlan P. Lindoy}
\author{Deep Lall}
\author{Sebastian~E.~de~Graaf}
\author{Tobias Lindstr\"om}
\affiliation{National  Physical  Laboratory,  Teddington,  TW11  0LW,  United  Kingdom}
\author{Ivan Rungger}
\email{ivan.rungger@npl.co.uk}
\affiliation{National  Physical  Laboratory,  Teddington,  TW11  0LW,  United  Kingdom}
\affiliation{Department of Computer Science, Royal Holloway, University of London, Egham, TW20 0EX, United Kingdom}

\begin{abstract}
       Qubit noise and fluctuations of the noise over time are key factors limiting the performance of quantum computers. Characterising them with high temporal resolution is challenging due to multiple overlapping stochastic processes such as discrete jumps and continuous drifts. Hence, experiments typically probe individual sources of fluctuations rather than concurrent fluctuations caused by multiple sources. To overcome this limitation we develop a framework comprising a noise characterisation method with minimal measurements allowing high temporal resolution, combined with a hierarchical discrete fluctuation auto-segmentation tool to disentangle the overlapping fluctuations without human intervention, enabling their characterisation and tracking over long times. We show that on transmon qubits the method can track and disentangle qubit frequency fluctuations with temporal resolution of a few tens of milliseconds over hours. This enables us to identify the origins of the fluctuations as overlapping charge parity and two-level-systems switching. Beyond insights into the fluctuation origins, our method also provides information that can be used to improve qubit calibration, error mitigation and error correction.
\end{abstract}    

\maketitle

\section{Introduction}
\label{sec_introduction}
The progress towards large-scale, fault-tolerant quantum computers has been driven in the recent decade by an increase in the number and quality of qubits within different hardware platforms~\cite{bluvstein2024logical,acharya2024quantum,reichardt2024demonstration,slussarenko2019photonic,RevModPhys.95.025003}. In superconducting qubits~\cite{kjaergaard2020superconducting,krantz_quantum_2019,siddiqi2021engineering} typical coherence times have increased from $1-10~\mu\mathrm{s}$ in the late 2000's~\cite{kjaergaard2020superconducting} to $>100~\mu\mathrm{s}$ coherence times available in publicly accessible devices today~\cite{kim2023evidence}, and even greater than $1~\mathrm{ms}$ coherence times in certain qubits~\cite{ganjam2024surpassing,somoroff2023millisecond}. Fidelities of operations may already be below the quantum error correction (QEC) threshold in some instances~\cite{bluvstein2024logical,acharya2024quantum,brock2024quantum,reichardt2024demonstration}, but further improvements to qubit quality are still needed to reduce the overheads of QEC and make fault-tolerant quantum computing for large scale practical applications feasible. In the context of superconducting qubit devices, various different strategies are being tested and implemented towards this aim, such as the development of novel qubits \cite{Nguyen2019HighCoherence}, improvement of materials~\cite{murray_material_2021,ganjam2024surpassing,Richardson_Lordi_Misra_Shabani_2020,siddiqi2021engineering,Oliver_Welander_2013}, and novel engineering solutions~\cite{Chistolini2024performanceof,mcewen_resisting_2024,chen2024phononengineering,Bargerbos2023Mitigation}. In each of these avenues it is essential to have a thorough understanding of the sources of noise present in the qubits in order to guide the development of solutions to reduce errors.

When modelling noise in individual qubits, the most commonly used metrics to estimate the quality of the qubits are the $T_1$ and $T_2$ times, which measure the qubit relaxation and dephasing rates, respectively~\cite{lall2025review}. However, additional noise contributions need to be accounted for in order to accurately describe qubit behaviour, including coherent errors such as those due to miscalibrations~\cite{vepsalainen2022improving,lucero2010reduced,blume2017demonstration}, non-Markovian noise~\cite{agarwal2023modelling,young_diagnosing_2020,shirizly_dissipative_2023,Prakash2024Characterizing,oda_sparse_2024,zhang_predicting_2021}, and errors introduced by the application of gates~\cite{blume2017demonstration}. It is also important to account for fluctuations in these noise contributions over time: fluctuations in noise at timescales that are longer than individual circuit durations but shorter than total experiment duration extending over a large number of circuit executions can manifest as non-Markovian noise in the qubits~\cite{agarwal2023modelling}. Since non-Markovian noise can also have other sources, such as coupling to near-resonant two-level-system (TLS) defects~\cite{muller2019towards}, characterisation of noise fluctuations at the timescales of a few circuit executions can contribute to identifying the physical origins of the noise, one of the main open questions in the field. In general, characterising noise fluctuations at varying timescales also allows for a better understanding of the device physics and stability over time, which can be an important factor in device performance~\cite{lall2025review,dasgupta2021stability, proctor2020detecting,dasgupta2020characterizing,yeter2022measuring,van2023evaluating,dasgupta2024stability,alam2019addressing,carroll2022dynamics, wan2019quantum,merkel2019magnetic, schlor2019correlating,de2021quantifying,wei2022measurement} and can significantly affect the complexity of quantum error mitigation and quantum error correction~\cite{kim2023evidence,acharya2024quantum}. 

In superconducting qubits, repeated measurements of qubit coherence times as well as qubit frequencies over timescales ranging from minutes to days have revealed significant change over time of these parameters~\cite{schlor2019correlating,burnett2019decoherence,Klimov_2018,muller2015interacting,yang_locating_2023,kono_mechanically_2024}.
Fluctuations in the qubit frequencies have been found to arise from a number of different sources, such as charge parity (CP) switching~\cite{riste_millisecond_2013,catelani_relaxation_2011,koch_charge-insensitive_2007,christensen_anomalous_2019,serniak_direct_2019,serniak_hot_2018,martinez_noise-specific_2023,diamond_distinguishing_2022,pan_engineering_2022,liu_observation_2024,papic_charge-parity_2024,mannila_superconductor_2022-1,liu_quasiparticle_2024,tennant_low-frequency_2022,gordon_environmental_2022,erlandsson_2023_parity}, jumps in the charge offsets~\cite{riste_millisecond_2013,thorbeck_tls_2022,wilen_correlated_2021,tomonaga_quasiparticle_2021}, coupling to TLS~\cite{schlor2019correlating,de_graaf_two-level_2020,paik_observation_2011,levine_demonstrating_2024}, and flux noise~\cite{vepsalainen2022improving}. Evaluating properties of the fluctuations such as their magnitudes and rates can help identify which of these physical sources is affecting a particular device. However, this requires characterisation of the fluctuations occurring at fast timescales.

Typically, methods for fast characterisation of noise effects involve designing the experiments to characterise only specific sources of noise. If a property is known to fluctuate between discrete states, for example due to CP switching or low-energy TLS, the circuits run on the qubits can be designed to simply choose between the known discrete states~\cite{riste_millisecond_2013,liu_observation_2024}.
If the functional form of the noise being characterised is known, Bayesian methods, which update the circuits on-the-fly to maximise information gain about the noise, can be used ~\cite{stenberg2014efficient,stenberg2016simultaneous,bejanin2021resonant}. If fast detection of events such as sudden drop in the $T_1$ time of the qubit need to be observed, circuits can be designed to only measure those specific noise effects~\cite{mcewen2022resolving,mcewen_resisting_2024,de2024evaluating}. Although these characterisation methods can allow observing specific effects at fast timescales, the main limitation is that they fail when the underlying assumptions about the expected form of the effect being observed are not valid. For example, this is the case when the discrete states drift or fluctuate themselves over the considered time-scales. Another limitation is that to evaluate correlations between different sources of fluctuations, different methods must be interleaved with each other~\cite{thorbeck_tls_2022}, which reduces the temporal resolution.

When multiple sources of fluctuations occur concurrently, it is challenging to characterise the different fluctuations and evaluate properties such as their rates and magnitudes. CP switching and coupling to TLS are known sources of qubit frequency fluctuations and both have been studied extensively independently~\cite{riste_millisecond_2013,christensen_anomalous_2019,serniak_direct_2019,serniak_hot_2018,schlor2019correlating,de_graaf_two-level_2020,paik_observation_2011}. To our knowledge, only Ref.~\cite{liu_observation_2024} analyses the concurrent fluctuations caused by these two sources, but in that case the fluctuation caused by the TLS is significantly larger than the one caused by CP switching, making it straightforward to disentangle the two types of fluctuations. In the general case, when the magnitudes of the different fluctuations are similar and the fluctuations overlap, disentangling these fluctuations is challenging, so that bespoke time-series processing methods must be developed and used.
A method that is used to detect fluctuations of a quantity between different discrete values is the hidden Markov model (HMM)~\cite{rabiner1986introduction}, in which a time series of observed values depends on a hidden state at each time, and where the hidden states are described by a single Markov chain. In the presence of multiple concurrent sources of fluctuations one has to include multiple stochastic processes, so that using a conventional HMM to describe the combined fluctuations can become computationally infeasible. In these cases, methods involving hierarchical HMMs~\cite{fine1998hierarchical} or multi-level random telegraph noise~\cite{Puglisi_Pavan_2014}, which model a time series as a composite hierarchy of multiple stochastic process, can be used. However, when one of the sources of fluctuation involves jumps between a range of continuous values rather than jumps between a few discrete values, representing it as an HMM can require a large number of hidden states, making the characterisation of the composite fluctuations impractical. Therefore in these instances, bespoke methods that can account for continuous-range fluctuations without an increase in model complexity must be developed.

Here we present a method overcoming these limitations for fast characterisation, building on gate execution sequences for idle qubit noise characterisation~\cite{agarwal2023modelling}, which allows for high temporal resolution by using minimal experimental data to obtain the noise model at short time intervals. Characterising and disentangling fluctuations of the noise requires a hierarchical fluctuation model that allows for fluctuations between a continuous range rather than just a few discrete values, as well as a tool to automatically process large amounts of fast-tracking data and map it to the fluctuation model. To address these needs, we develop the hierarchical discrete fluctuation auto-segmentation (HDFA) method, which builds on techniques used in hierarchical HMMs and tools from time-series segmentation~\cite{aminikhanghahi2017survey,truong2020selective} to map concurrent fluctuations to a composite of Markov chains as well as arbitrary piecewise-continuous stochastic processes. By mapping the fluctuations to a composite of stochastic processes, the method thus enables disentangling different sources of fluctuations and allows their characterisation.
We thereby obtain a framework for characterising noise fluctuations in qubits consisting of the detection of noise fluctuations at fast timescales, disentanglement of concurrent fluctuations and their independent characterisation, and finally utilisation of the disentangled fluctuations to tailor quantum algorithms, develop tools to mitigate the effects of fluctuations, or gain insights into the physical origins of the fluctuations. We demonstrate the capability of this framework by applying it to a superconducting qubit device and characterising the persistent fluctuations present with a resolution of a few tens of milliseconds over multiple hours. We find that concurrent fluctuations are due to CP switching on very fast timescales, combined with jumps in the qubit charge offset as well as coupling to charge-dipole TLS at slower timescales.

\section{Methodology}
\label{sec:methodology}

\begin{figure*}
    \centering
    \includegraphics[width=\textwidth]{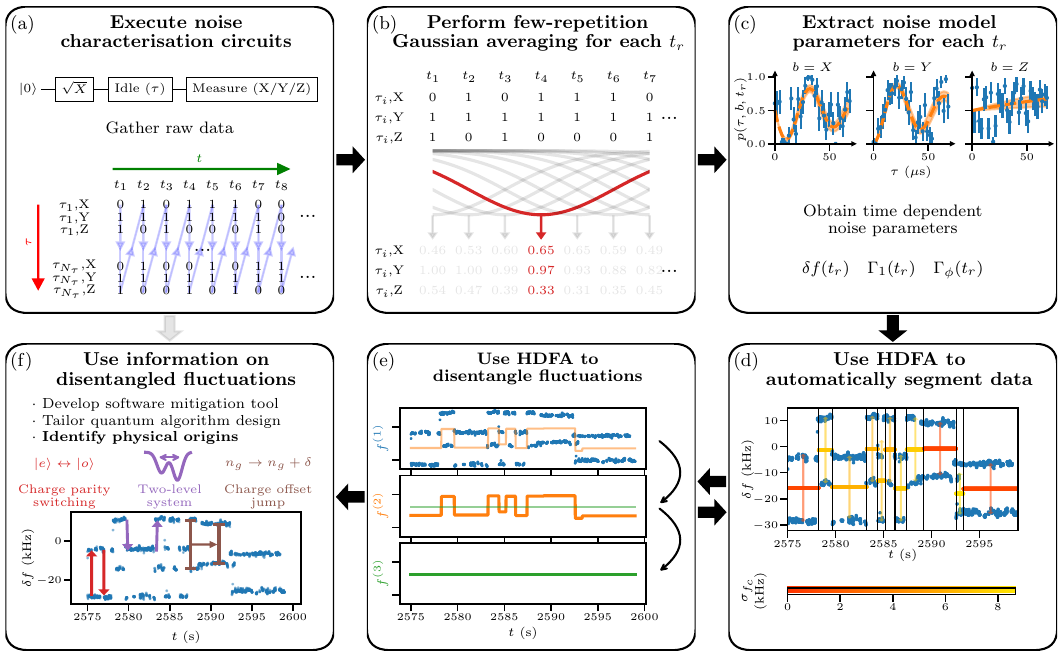}
    \caption{Schematic representing the framework for characterising and disentangling noise fluctuations. 
    (a) The single-qubit circuit for the noise characterisation is shown, containing a preparation of the $\ket{+i}$ state via application of the $\sqrt{X}$ gate, followed by delay of time $\tau$, and then finally a measurement in one of the $X$, $Y$, and $Z$ bases, which includes a basis change operation for the $\{ X,Y\}$ bases before measurement in the computational basis. Circuits for different idle times $\tau$ and measurement basis are repeated cyclically and each repetition is indexed by $t_r$ which corresponds to the start time of that sequence repetition. The result of each circuit is a binary outcome $0/1$ corresponding to the measurement result. 
    (b) Once the data has been gathered, the binary results are averaged using a Gaussian moving window to estimate the probability of obtaining $0$ for each different circuit at each time step $t_r$. 
    (c) For each $t_r$, a noise model including frequency detuning, relaxation, and pure dephasing is fitted to the data using non-linear regression to obtain the corresponding noise parameters $\delta f, \Gamma_1, \Gamma_\phi$. The plots in (c) show an example of Gaussian averaged data (blue) and the fitted noise model prediction (orange). The error bars denote the standard error of the mean due to averaging over very few repetitions.
    (d) Discrete fluctuations in $\delta f(t)$ are characterised by using the developed HDFA tool to map the data to the model $\delta f = f_c \pm f_\Delta/2$. The blue dots in the panel denote the evaluated $\delta f$ from (c) and the solid lines denote $f_c$ and $f_\Delta$. The colours of the solid lines represent the evaluated uncertainty in $f_c$. Along with fast fluctuations of magnitude $f_\Delta$, fluctuations in $f_c$ and $f_\Delta$ themselves can also be seen.
    (e) The HDFA tool is used to disentangle fluctuations by iteratively extracting out the fastest fluctuations from the data. After removing the fastest fluctuations in $f^{(1)} = \delta f$ (blue), there are still remaining fluctuations in $f^{(2)} = f_c$ (orange). The fastest fluctuations in $f^{(2)}$ are then removed to obtain $f^{(3)}$ (green), which is constant for this period of data and shows successful disentanglement of the different sources of discrete fluctuations.
    (f) Finally, the properties of fluctuations at each timescale can be used towards multiple aims, for example, mapping to compatible physical models of the origins of the fluctuations. In this example, the fastest fluctuations are attributed to charge-parity switching (illustrated by red arrows), fluctuations in their amplitude $f_\delta$ are attributed to charge offset jumps (illustrated by brown arrow and lines), and fluctuations in the centre frequency $f_c$ are attributed to coupling to TLS (illustrated by purple arrows).}
    \label{fig:schematic}
\end{figure*}    

Our framework for the characterisation of different sources of noise fluctuations is summarised in \figref{fig:schematic}.
It consists of two parts: in the first part a set of circuits and measurements to characterize the properties of the qubit at different times $t$, including the qubit frequency, is executed (\figref{fig:schematic}a-c); in the second part an automated approach is introduced to disentangle the frequency fluctuations over time and to assign them to potential physical origins (\figref{fig:schematic}d-f).
Below we provide the methodologies involved in each step of the framework.

\subsection{Noise characterisation circuits}
\label{subsec:noise_characterisation_circuits}
The quantum circuits that we execute to characterise the noise are motivated by idle qubit tomography experiments used to measure noise in idle qubits~\cite{agarwal2023modelling,blume-kohoutVolumetricFrameworkQuantum2020,zhang_predicting_2021,tripathi2022suppression}. Starting from the initial $\ket{0}$ state, one first applies a $\pi/2$ rotation about the $X$ axis to prepare the qubit in the $\ket{\psi} = \frac{\ket{0} - i\ket{1}}{\sqrt{2}}$ state. Then the qubit is left idle for a time $\tau$. Finally, a measurement in one of $\{ X,Y,Z\}$ bases is performed, where an appropriate basis change operation is applied before measurement in the computational basis for the $\{ X,Y\}$ bases. These quantum circuits are depicted in~\figref{fig:schematic}(a). The expectation values of the Pauli operators are given by $\langle\hat{\sigma}_{\alpha}(\tau)\rangle = \mathrm{Tr}[\hat{\rho}(\tau)\hat{\sigma}_{\alpha}]$, with $\alpha\in\{X,Y,Z\}$,  and where $\hat{\sigma}_{\alpha}$ are the Pauli operators and $\hat{\rho}(\tau)$ is the single qubit density matrix after the idle evolution. Then the aggregated measurements in the different bases correspond to performing single qubit state tomography~\cite{nielsenGateSetTomography2021,leibfried1996experimental,banaszek2013focus,blume2010optimal}, and allow one to obtain the expectation values $\expval{\sx(\tau)}, \expval{\sy(\tau)}$, and $ \expval{\sz(\tau)}$, which in turn allow estimating the single qubit density operator $\hat{\rho}(\tau)$. 

Circuits for different values of idle time $\tau \in \{\tau_1,\tau_2,\ldots,\tau_{N_\tau}\}$ and measurement basis $b \in \{X,Y,Z\}$ are executed repeatedly in a single experiment with the ordering as follows. First, a single execution of the circuit with $\tau = \tau_1$ and measurement basis $b = X$ is performed. Next, the circuit for $\tau = \tau_1$ and $b=Y$ is run, followed by $\tau = \tau_1$ and $b=Z$. The next three circuits correspond to $\tau = \tau_2$ and $b=X$, $b=Y$, and $b=Z$. This continues until the circuit for the last value of $\tau$ and the last measurement basis, i.e. $\tau = \tau_{N_\tau}$ and $b=Z$, is run. We denote this as a single sequence of circuit executions, which includes a number of circuits, $N_c$, equal to $N_c=3 N_\tau$, where the factor of 3 is due to the three measurement bases. The total time to execute this set of circuits, $\delta t$, then is $\delta t = 3\sum_{i=1}^{N_\tau} (\tau_i+t_\mathrm{other})$, where $t_\mathrm{other}$ includes times for the state preparation and basis change gates, qubit measurement and reset, as well as other contributions such as delays added between the execution of circuits.

Since each measurement can only give either 0 or 1 as output, to obtain the typically non-integer expectation values $\expval{\sx(\tau)}, \expval{\sy(\tau)}$, and $ \expval{\sz(\tau)}$ one needs to repeat the execution of the above sequence of circuits multiple times and average over the results. We denote the number of sequence repetitions as $N_s$, and the integer index of a repetition as $r$. The time at which the first circuit in each repetition with index $r$ is executed is denoted as $t_r$, where we set $t_1 = 0$. The circuits and the ordering are shown in~\figref{fig:schematic}(a). The total time to obtain the expectation values then is $\Delta_t=N_s\; \delta t$. Note that $\Delta_t$ is usually orders of magnitude larger than $\tau_{N_\tau}$, since a large number of circuits is typically run within the time $\Delta_t$. If the exact pulse sequence executed on the hardware is known, it can be directly used to obtain $\Delta_t$. Otherwise, $\Delta_t$ can be approximated by using the total time to run the experiment and assuming that each repetition takes an equal amount of time.

To be able to track fluctuations of qubit properties over fast time scales one needs to minimize $\Delta_t$, and hence both $N_s$ and $\delta t$. However, reducing the values of these quantities increases the statistical uncertainty due to reduced number of samples used in the averaging process, and therefore has the drawback that it increases the uncertainty in the obtained expectation values and hence in qubit properties. One therefore needs to choose $N_s$ and $\delta t$ aiming to achieve an optimal trade-off that minimizes $\Delta_t$ while also keeping the uncertainties in the extracted qubit properties small. 

\subsection{Gaussian averaging}
\label{subsec:gaussian_averaging}
We denote the binary outcome of the circuit corresponding to idle duration $\tau$, measurement basis $b$, and repetition $r$ as $B(\tau,b,t_r)$, such that $B=1$ if the qubit is successfully measured in the $\ket{0}$ state that it started with. The value of $B(\tau,b,t_r)$ corresponds to a particular random output sampled from the probability of obtaining $0$ from the probability distribution at each time $t_r$, denoted by $p^*(\tau,b,t_r)$. As outlined above, to obtain the estimates of the expectation values at a time $t_r$ one may average the measured values over $N_s$ repetitions of the set of circuits executed at times {$t_r$, $t_{r+1}$, \dots, $t_{r+N_s-1}$. This effectively gives the expectation values as running average with length $N_s$. 
Rather than using a fixed window size, one can instead use a Gaussian moving average~\cite{smith1997scientist}, where results further away in time from $t_r$ are given smaller weights when evaluating an estimate for $p^*(\tau,b,t_r)$. To distinguish the true probability distribution from the probability distribution estimated by the experimental results, we use $p^*(\tau,b,t_r)$ to denote the true distribution and $p(\tau,b,t_r)$ to denote the estimated distribution. With the Gaussian moving average, the estimated probability is given by
\begin{equation}
    p(\tau,b,t_r) = \frac{\sum_{i} {B(\tau,b,t_i)}w(i,r)}{\sum_{i} {w(i,r)}},
    \label{eq:gaussian_moving_average_def}
\end{equation}    
where $w(i,r)$ is the weight assigned to the particular contribution, and is given by
\begin{equation}
    w(i,r) = e^{-\frac{(i-r)^2}{2W_G^2}},
    \label{eq:gaussian_averaging_weight_definition}
\end{equation}    
with $W_G$ corresponding to the width of the Gaussian kernel used for the averaging. This is visually depicted in~\figref{fig:schematic}(b). Larger values of $W_G$ increase statistical precision due to effectively averaging over more data, but reduce temporal resolution due to averaging over features only occurring at fast timescales. As given by the standard quantum limit~\cite{vittorio2004quantum}, the uncertainty in the expectation value due to a finite number of repetitions scales as $W_G^{-1/2}$. Due to Gaussian averaging, the smallest time interval over which differences in $p(\tau,b,t_r)$ can be resolved is approximately $\delta t\;W_G$. The temporal resolution due to Gaussian averaging is therefore given by the inverse of this quantity. To maximise temporal resolution for performing fast-tracking, the value of $W_G$ must be set to be the smallest value for which fluctuations in the noise parameters can be sufficiently detected for a given set of data. We denote the averaging using such a small window as the few-repetition averaging protocol. The optimal value of $W_G$ generally differs for each noise parameter, since the number of measurements required to obtain a given target relative uncertainty depends on the magnitude of the fluctuations of that parameter and on the sensitivity of the measured quantities to that parameter. In this work, we primarily target tracking qubit frequency fluctuations, therefore $W_G$ is optimised for tracking fluctuations in the qubit frequency.

In~\appref{app:sec:gaussian_window_averaging} we compare the Gaussian averaging procedure with a fixed window moving average and find that for $W_G<5$ the Gaussian averaging can lead to more precise estimates of noise parameters while still having a similar temporal resolution. Thus, the Gaussian averaging protocol is used in this work.

\subsection{Noise parameters evaluation}
\label{subsec:noise_model_evaluation}
It has been shown that a noise model consisting of qubit relaxation, dephasing, detuning of the calibrated qubit driving frequency from the true frequency, and non-Markovian noise can represent noise in idle Transmon qubits  accurately~\cite{agarwal2023modelling,shirizly_dissipative_2023,zhang_predicting_2021}. Fluctuations in these noise parameters at different timescales show up as different effects on the averaged data. For example, frequency fluctuations at timescales faster than a single circuit execution show up as dephasing noise~\cite{krantz_quantum_2019}, frequency fluctuations at timescales longer than individual circuits but shorter than the averaging time show up as non-Markovian noise~\cite{agarwal2023modelling}, and fluctuations at timescales longer than averaging length show up as fluctuations of the averaged data~\cite{schlor2019correlating}.
If the only source of non-Markovianity are fluctuations in qubit or noise properties, then at time-scales shorter than the fluctuations in noise parameters the noise is approximately Markovian.
In this case, and given that with our few-repetition averaging protocol we effectively average only over short times, the qubit frequency is approximately constant for most of the evaluated times $t_r$, except when a frequency jumps occurs within the $W_G$ window. For the majority of times therefore a Markovian noise model captures the qubit behaviour well. We hence use the noise model of Ref.~\cite{agarwal2023modelling}, but without the non-Markovian contributions.
It should be noted that if the true source of non-Markovian noise is coherent interaction with a defect rather than frequency fluctuations, then one has to use the full noise model of Ref.~\cite{agarwal2023modelling}. 

For completeness, the noise model is described below (for details see Ref.~\cite{agarwal2023modelling}.):
\begin{enumerate} [noitemsep,topsep=0pt]
    \item $\delta f$: this is the detuning of the qubit frequency from the calibrated value, and is given by $\delta f=f-f_0$, where $f$ is the true qubit frequency and $f_0$ is the calibrated qubit driving frequency. This calibration error introduces an additional term in the Hamiltonian of the form $\frac{2\pi\delta f}{2} \sz$, where $\sz$ is the Pauli-z operator.
    \item $\Gamma_{1}$: this is the qubit relaxation rate. It is related to the $T_1$ time of the qubit via $\Gamma_{1} = \frac{1}{T_1}$~\cite{krantz_quantum_2019}. This noise is modelled in the Lindblad Master equation formalism as $\dot{\hat{\rho}} = \hat{L}_1 \hat{\rho} \hat{L}_1^\dagger - \frac{1}{2}\{\hat{L}_1^\dagger\hat{L}_1,\hat{\rho}\}$ with $\hat{L}_1 = \sqrt{\Gamma_1}\hat{\sigma}_{-}$, where $\hat{\sigma}_{-}$ is the lowering operator.
    \item $\Gamma_{\phi}$: this is the pure dephasing rate. It is related to the $T_2$ time of the qubit via $\Gamma_{\phi} = \frac{1}{T_2} - \frac{1}{2T_1}$~\cite{krantz_quantum_2019}. This noise is modelled in the Lindblad Master equation formalism as $\dot{\hat{\rho}} = \hat{L}_\phi \hat{\rho} \hat{L}_\phi^\dagger - \frac{1}{2}\{\hat{L}_\phi^\dagger\hat{L}_\phi,\hat{\rho}\}$ with $\hat{L}_\phi = \sqrt{\frac{\Gamma_\phi}{2}} \sz$.
\end{enumerate}       

If the noise parameters do not change over time one has a time-independent probability $p^*(\tau,b,t_r) = p^*(\tau,b)$, with\begin{equation}
    p^*(\tau,b)  =         
\begin{cases}
    \frac{1-e^{-(\frac{\Gamma_{1}}{2}+ \Gamma_{\phi}) \tau}\sin (2\pi \delta f \tau)}{2}  & \text{if $b = X$,} \\
    \frac{1-e^{-(\frac{\Gamma_{1}}{2}+ \Gamma_{\phi}) \tau}\cos (2\pi \delta f \tau)}{2}  & \text{if $b = Y$,} \\
    1-\frac{e^{-\Gamma_{1} \tau}}{2} & \text{if $b = Z$.} \\
\end{cases}    
\end{equation}
These equations show that in this case the $X$ and $Y$ measurement basis probabilities are described by a single-frequency decaying oscillation, while the $Z$ basis probability exhibits an exponential decay with $\tau$. 
When evaluating the noise model parameters we assume that the noise is constant within each averaging window but can change between consecutive times, i.e. between $t_r$ and $t_{r+1}$, and hence use the above equations to evaluate the noise parameters for all values of $t_r$ to obtain $\delta f(t_r)$, $\Gamma_1(t_r)$, and $\Gamma_\phi(t_r)$.
Non-linear regression with a differential evolution optimizer~\cite{storn1997differential,scipy} is used to obtain these noise parameters from $p(\tau,b,t_r)$. An illustrative example of the procedure is shown in~\figref{fig:schematic}(c). 

When analysing fluctuations in the noise parameters over time, it is important to have quantified uncertainties of the estimated noise parameters, since that can allow distinguishing true fluctuations which are much larger than the uncertainties from apparent fluctuations which arise due to the uncertainties in the noise parameter estimates. To evaluate the uncertainties in the noise parameters, a bootstrap method is used that includes uncertainty due to imperfect model fitting as well as uncertainty due to the finite number of measurement samples. The details are provided in~\appref{app:sec:frequency_uncertainty_estimation}.

\subsection{The hierarchical discrete fluctuation auto-segmentation algorithm}
\label{sec:hdfa}
This section presents our HDFA algorithm that we developed to disentangle and characterise multiple concurrent discrete fluctuations. Consider the time series of a variable $f(t_r)$, where $f(t_r)\in \mathbb{R}$. Suppose $f(t_r)$ exhibits only discrete jumps between two values, corresponding to a stochastic process described by a random telegraph noise (RTN)~\cite{paladino_1f_2014}. Then, $f(t_r)$ can be written as
\begin{equation}
    f(t_r) = f_c + s(t_r)\frac{f_\Delta }{2},
\end{equation}    
where $f_c$ denotes the centre of the two RTN states, $f_\Delta$ is the RTN fluctuation magnitude, and $s(t_r) \in \{-1,1\}$ is the RTN state determined by an underlying stochastic process. The task of modelling these fluctuations then requires obtaining the model parameters $f_c, f_\Delta$, and $s(t_r)$ from the given $f(t_r)$. One approach to obtain these parameters is to use a Gaussian hidden Markov model (HMM)~\cite{rabiner1986introduction}, in which the unknown - or hidden - states of the HMM are $s(t_r)$ and the  term ``Gaussian" refers to $f(t_r)$ being described by two Gaussian distributions centred around $f_c \pm \frac{1}{2} f_\Delta$, which account for broadened peaks when $f(t_r)$ is obtained from experiments. In this specific case of a single RTN, to obtain $f_c$ and $f_\Delta$ the Baum-Welch algorithm~\cite{rabiner1986introduction} can be used, which involves iteratively updating the HMM parameters and maximising the likelihood of the observed data given the model parameters. Then, the Viterbi algorithm~\cite{rabiner1986introduction} can be used to find the most likely sequence of hidden states $s(t_r)$.

In the more general case where multiple physical sources of fluctuations or drifts are present in a device, a single RTN cannot describe all fluctuations in a noise parameter. In this case, $f_c$ and $f_\Delta$ are no longer constant over time, and instead can exhibit continuous drifts and discrete jumps with time. If these parameters only exhibit discrete jumps between a few states, their time dependence can be accounted for by introducing hidden states which govern their jumps with time. However, in the general case where $f_c$ and $f_\Delta$ exhibit fluctuations between many discrete values or even over a continuous range, a large number of hidden states would be needed to discretize the continuous fluctuations. Such a model is challenging to fit to the data, and leads to a lack of interpretability of the fluctuations. To overcome this, a model and method is needed, which provides structure to fluctuations of $f_c$ and $f_\Delta$, and which can then allow obtaining time-dependent HMM parameters. This is achieved by the HDFA algorithm developed in this work.

The HDFA algorithm assumes that $f(t_r)$ exhibits discrete jumps which are described by a hierarchy of stochastic processes, and which affect $f(t_r)$ at different timescales. The levels of hierarchy are denoted as $\mathcal{S}^{(n)}$, where $n$ is the hierarchy level. We consider the case where each stochastic process is described by an RTN, although the method can be generalised to allow for stochastic processes showing discrete jumps between more than two states. Let $f^{(1)}(t_r) = f(t_r)$ correspond to fluctuations at the hierarchy level $\mathcal{S}^{(1)}$, $f^{(2)}(t_r)$ at the level $\mathcal{S}^{(2)}$, and so on, where increasing levels of the hierarchy correspond to stochastic processes at slower timescales. Since each stochastic process is a RTN, the variable $f^{(n)}(t_r)$ can be written as
\begin{equation}
    f^{(n)}(t_r) = f_c^{(n)}(t_r) + s^{(n)}(t_r)\frac{f_\Delta ^{(n)}(t_r)}{2},
\end{equation}    
where the centre and magnitude of the RTN are now also time-dependent.
Although $f_\Delta^{(n)}(t_r)$ and $f_c^{(n)}(t_r)$ can both themselves exhibit a hierarchy of fluctuations, in our model we let $f_\Delta^{(n)}(t_r)$ be arbitrary piecewise-constant while $f_c^{(n)}(t_r)$ is described by the remaining hierarchy of stochastic processes affecting $f(t_r)$, such that the centre of the RTN at a particular level of hierarchy is described by the composite of fluctuations at all slower levels of hierarchy. Thus, the recursive relation defining the hierarchy of fluctuations is
\begin{equation}
f^{(n+1)} = f_c^{(n)}.
\end{equation}
By construction, $f^{(n)}$ has fluctuations at progressively slower timescales for increasing $n$, since $f^{(n+1)}=f_c^{(n)}$ is obtained after removing the fastest fluctuations from $f^{(n)}$. Thus, larger values of $n$ correspond to fluctuations at successively slower timescales. This hierarchy of fluctuations is visually depicted in~\figref{fig:schematic}(e).

We now discuss how time series data can be mapped to this hierarchical model of fluctuations. In time series processing, the task of evaluating when the probability distribution of a stochastic process changes is known as change point detection or time series segmentation~\cite{aminikhanghahi2017survey,truong2020selective}. Typically, this is performed when the underlying stochastic process is a continuous random variable and not a discrete random variable as in the case of RTNs. However, as we show below, the time series segmentation can also be performed to segment data into a hierarchy of fluctuations, each corresponding to discrete stochastic jumps. 

HDFA involves recursively obtaining $f_c^{(n)}, s^{(n)}$, and $f_\Delta ^{(n)}$ from $f^{(n)}$. Since $f_c^{(n)}$, $f_\Delta^{(n)}$ exhibit slower, piecewise-constant behaviour while $s^{(n)}$ exhibits the faster timescale RTN, the task at each step of the recursion is to segment the data such that each segment corresponds to a constant value of $f_c^{(n)}$ and $f_\Delta^{(n)}$, while still containing the faster timescale fluctuations of $s^{(n)}$. 
To perform this segmentation, $f^{(n)}(t_r)$ is iterated over sequentially increasing $t_r$ at each step. Let $\mathcal{T}(t_r)$ represent all time steps which are part of the same segment as $t_r$. At each step, the new value of $f^{(n)}(t_r)$ is added to the list of values of $f^{(n)}$ in the previous segment, i.e. $f^{(n)}(t \in \mathcal{T}(t_{r-1}))$, and the Baum-Welch algorithm is run to fit an HMM to the augmented segment. Then, the mean-log-likelihood, which is the mean of the base-10 logarithm of the probability of obtaining each observed $f^{(n)}$ in the augmented segment given the HMM, is computed. If a fluctuation of $f_c^{(n)}$ or $f_\Delta^{(n)}$ has occurred between $t_{r-1}$ and $t_r$, then in general $f^{(n)}(t_r)$ is different from the two discrete values describing the HMM of the un-augmented segment. This results in a reduction of the mean-log-likelihood, and for sufficiently significant fluctuations the mean-log-likelihood will fall below a pre-set threshold $\lambdall$. If a fluctuation of $f_c^{(n)}$ or $f_\Delta^{(n)}$ has not occurred, the addition of the new data point to the segment will not significantly change the mean-log-likelihood. Thus, if the mean-log-likelihood is above the threshold $\lambdall$ then $t_r$ is appended to the prior segment $\mathcal{T}(t_{r-1})$, while if it is below the threshold, the prior segment is terminated and a new segment is started with $t_r$. Upon iteration over all $t_r$, the time series of $f^{(n)}$ is fully segmented. An example of the segmentation procedure is shown in~\figref{fig:schematic}(d). Noise in the time series, such as due to uncertainty in the estimation of $f^{(n)}$, can lead to segments terminating early due to low mean-log-likelihoods for small segments. To overcome this, a minimum length of each segment, $L_\mathrm{min}$, is imposed to avoid early termination of HMM segments.

The hyperparameters $\lmin$ and $\lambdall$ both set the limits of the auto-segmentation procedure. The parameter $\lmin$ bounds the minimum timescale of $f_c (t)$ and $f_\Delta (t)$ fluctuations that can be detected, since fluctuations occurring faster than the $\lmin$ can be missed due to the minimum required segment length. The parameter $\lambdall$ limits the minimum change in $f_c$ or $f_\Delta$ that can be detected, because small changes in these properties, which do not sufficiently reduce the mean-log-likelihood, will not be detected. Thus, it is important to choose appropriate values for these hyperparameters. The method used for choosing these hyperparameters first involves running the segmentation for a range of hyperparameters. Then, the number of segments that the time series is divided into is plotted against the threshold $\lambdall$~\cite{haynes2017computationally}. The number of segments increase with larger thresholds, since the segment termination criteria is stricter. However, when the threshold becomes overly strict, the number of change points starts increasing at a faster rate. The value of the threshold at which this occurs is chosen as $\lambdall$~\cite{haynes2017computationally}. More details on the automated evaluation of these hyperparameters are provided in~\appref{app:subsec:automated_hyperparameter_evaluation}.

After the segmentation has been performed, $f_c^{(n)}$ and $f_\Delta^{(n)}$, as well as their uncertainties $\sigma_{f_c^{(n)}}$ and $\sigma_{f_\Delta^{(n)}}$ are evaluated for each segment to finally obtain the piecewise-constant $f_c^{(n)}(t_r)$ and $f_\Delta^{(n)}(t_r)$. Details on this evaluation, including how the uncertainties in $f^{(n)}$ are propagated to $f_c^{(n+1)}$ and $f_\Delta^{(n+1)}$, are provided in~\appref{app:subsec:auto-segmentation_uncertainty}. Properties of the fluctuation at that timescale, such as the transition rates between the two states denoted as $\nu^{(n)}_{1\rightarrow0}$ and $\nu^{(n)}_{0\rightarrow1}$, as well as the dynamics and distribution of $f_\Delta^{(n)}$, are then evaluated to characterise the fluctuation. The details on how the transition rates are evaluated while accounting for jumps that may be missed are presented in~\appref{app:subsec:hdfa_switching_rates}.

This segmentation is performed recursively for increasing $n$, until $f^{(n)}$ is constant or does not show a sufficient number of fluctuations over the considered time-range for reliable characterisation. At termination of the algorithm all stochastic processes in the hierarchy of fluctuations at different timescales have been disentangled from each other and characterised independently. An example of the recursive disentanglement of fluctuations is shown in~\figref{fig:schematic}(e).

\subsection{Mapping of fluctuation origins and optimising of control and application algorithms enabled by HDFA}
Having disentangled the different kinds of fluctuations present, the information extracted at each level of the fluctuation hierarchy includes their rates, their amplitudes, and the changes of these properties over time. This information can then be used to multiple aims such as tailoring the design of quantum algorithms, error correction or mitigation techniques, optimal control software for fluctuation mitigation, or gaining insights as to the physical origins of the fluctuations. As an example, in~\figref{fig:schematic}(f) we schematically present a set of potential physical origins associated with the various fluctuations. Below, we discuss how the information extracted from the fluctuation can be used towards these aims. 

To evaluate the potential physical origins of different fluctuations, one can compare the rates and magnitudes of the characterised fluctuations with the expected values for these parameters for different physical sources of fluctuations. 
For example, if the fluctuating noise source is due to CP switching, the fluctuation amplitude $f^{(n)}_\Delta$ can be expected to change with time, to have a continuous distribution, and to have a maximum in its distribution that is determined by the known design parameters of the qubit~\cite{riste_millisecond_2013}. Thus, by comparing the observed magnitude of fluctuations with the expected value based on the qubit properties, one can determine whether or not CP switching would be a compatible physical origin for the fluctuations. One can perform similar checks for other sources of discrete fluctuations such as coupling to TLS~\cite{schlor2019correlating,de_graaf_two-level_2020}.

Another example of how the properties of the disentangled fluctuations can be utilised is to build software tools to mitigate the fluctuations by changing the qubit gate parameters such as the pulse frequencies in real-time. Running the full noise characterisation circuits and doing the noise model fitting can be slow in this context. However, different levels of characterisation can be performed at different timescales to significantly reduce the overhead. For example, over longer timescales the detailed characterisation such as the one presented in this work can be performed and used to identify the dominant, persistent fluctuations. Then, before running quantum algorithms, a minimal characterisation can be performed and used to identify the different discrete values of the contributing noise sources at the timescales of the algorithm, and evaluate the minimal set of circuits that can distinguish between the different discrete values~\cite{liu_observation_2024}. Finally, in conjunction with the quantum algorithm, the minimal set of circuits can be run to evaluate the approximate noise model and update the gate calibrations and/or quantum circuits run as part of the quantum algorithm. Such a feedback-based tool can improve on the existing stabilisation methods ~\cite{vepsalainen2022improving} by accounting for the discrete nature of the observed fluctuations.

\section{Results}
\label{sec:results}
\subsection{Experimental setup}
\label{subsec:experimental_details}
We run experiments on the 5 transmon qubit \textit{ibmq\_lima} device~\cite{ibm_quantum}. Properties of this superconducting qubit device are provided in~\appref{app:subsec:device_details}. The protocol described in~\secref{subsec:noise_characterisation_circuits} is run on 3 non-neighbouring qubits, namely qubits 0, 2, and 4, in parallel. By using non-neighbouring qubits we minimize crosstalk effects while maximising the amount of data gained in each run. We verified that the crosstalk across these qubits is negligible compared to other noise sources in the idle qubit tomography experiments performed. The idle times $\tau$ are uniformly spaced with $N_\tau = 33, \tau_1 = 0~\mu\mathrm{s}$, and $\tau_{33} = 68.3~\mu\mathrm{s}$, which minimises $\sum_i\tau_i$ while still allowing for accurate characterisation of the expected range of noise parameters. Since we measure in three different bases for each idle time, the total number of unique circuits in a sequence is $N_c=3 N_\tau = 99$. In a single execution script that is run on the device, these are repeated for $N_s = 20,000$ repetitions. We submit $100$ such experiments to the queue, such that the total number of repetitions is $2\times10^6$
and the total number of circuit executions is $\sim 2\times10^8$. Most of these 100 experiments were executed consecutively on hardware, except for a long break of $\sim 8~\mathrm{h}$ after experiment 37. 
The time at which repetition $r$ within experiment $j$ was executed, is evaluated using the relation
\begin{equation}
t_{r}= (T_j^\mathrm{start}-T_0^\mathrm{start}) + (r-jN_s-1)  \frac{T_j^\mathrm{duration}}{N_s},
\end{equation}
where $T_j^\mathrm{start},T_j^\mathrm{duration}$ are the reported start time and duration for experiment $j$, respectively, and $r\in\{1,2,3,..,2\times10^6\}$. Note that for ease of notation in the remaining part of the manuscript the subscript $r$ is omitted for the time, so that only $t$ is used instead of $t_{r}$.
More details on the experimental runs are provided in~\appref{app:sec:experiment_device_details}.

\subsection{Noise model evaluation results}
\label{subsec:noise_model_evaluation_results}
\begin{figure}[thb!]
    \includegraphics[width=\columnwidth]{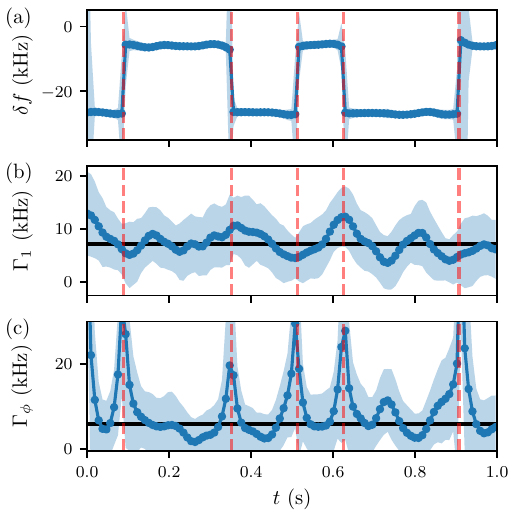}
    \caption{Example of the noise model evaluation results obtained for qubit 2. Panels (a), (b), and (c) show the noise parameters $\delta f, \Gamma_1,$ and $\Gamma_\phi$, respectively. The shaded region corresponds to the $95\%$ confidence interval obtained using the bootstrapping procedure. In (a), discrete fluctuations can be observed with the uncertainties smaller than the marker size for most times. In (b) and (c), the fluctuations are continuous and the results have a significant relative uncertainty. The horizontal black lines in (b) and (c) denote the weighted mean of the noise parameters, and largely lie within the confidence intervals. In (c), sudden increases in $\Gamma_\phi$, which are co-located with the frequency jumps in (a), can be seen, with the location of frequency jumps indicated by the vertical dashed red lines.}
    \label{fig:noise_model_short_time}
\end{figure}    
We first present results for the evaluation of the noise model parameters using the method presented in~\secref{subsec:noise_model_evaluation} and using the Gaussian averaging procedure introduced in~\secref{subsec:gaussian_averaging}. The smallest values of $W_G$ that still lead to sufficient accuracy in the estimates of $\delta f$ are found to be $W_G = 2,2,4$ for qubits $0,2,4$, respectively, and these are used to obtain the data in this manuscript. In~\appref{app:sec:gaussian_window_averaging} we show that the shortest duration of a $\delta f$ jump that can still be detected is $\sim 30,60~\mathrm{ms}$ for $W_G=2,4$, respectively, corresponding to a few tens of milliseconds temporal resolution for the fast-tracking.

In~\figref{fig:noise_model_short_time} the three noise model parameters for qubit 2 obtained for the first one second of the experiment are shown as a representative example, along with their $95\%$ confidence intervals. The confidence interval for $\delta f$ is estimated by $1.96\sigma_{\delta f}$, assuming a normal distribution of errors, where $\sigma_{\delta f}$ is the standard error obtained via the bootstrapping procedure described in \appref{app:sec:frequency_uncertainty_estimation}. The confidence intervals for $\Gamma_1$ and $\Gamma_\phi$ are obtained analogously. For $\delta f$ one can observe discrete fluctuations of magnitude $\sim 20\kHz$. These jumps in the parameters are significantly larger than the median confidence interval of $\sim1~\mathrm{kHz}$, which can be considered the frequency resolution limit of the model. Thus, this shows the presence of well defined discrete RTN-type frequency fluctuations in the noise at sub-second timescales. In the subsequent sections, these discrete fluctuations are characterised by using the HDFA algorithm presented in Sec. \ref{sec:hdfa}.

\begin{figure*}[tbh]
    \centering
    \includegraphics[width=\textwidth]{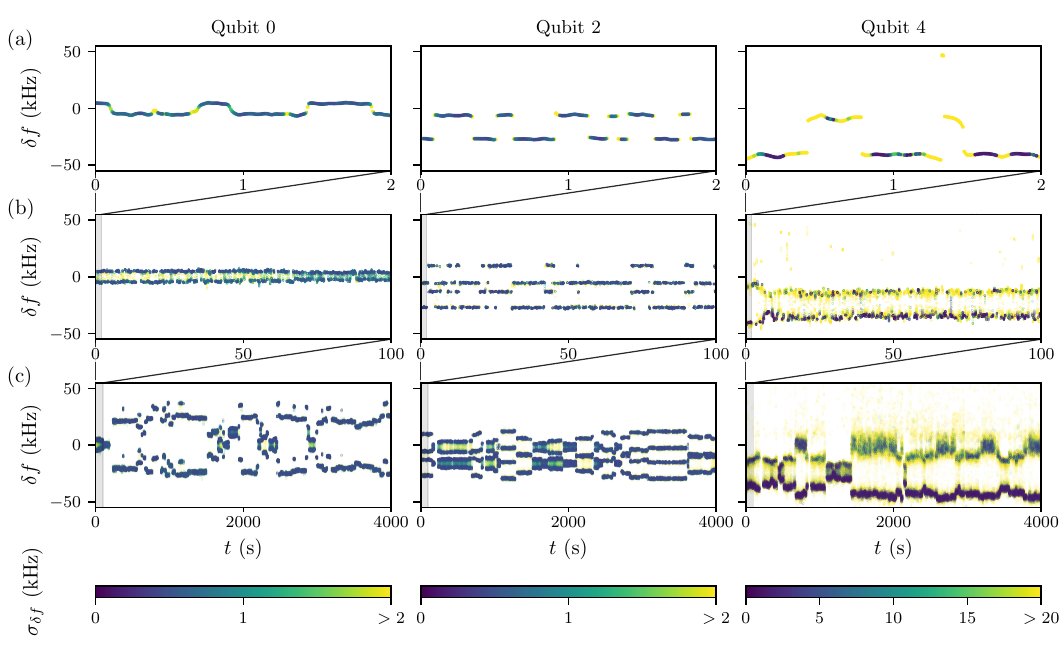}
    \caption{Results of the qubit frequency estimation procedure illustrating fluctuations at different timescales (shown in different rows), and for the three qubits (shown in different columns). The colours denote the uncertainty in the noise parameter estimate. At the $\sim 1~\mathrm{s}$ timescale (top row), large discrete jumps in $\delta f$ between two values occur. Over the $100~\mathrm{s}$ timescale (middle row), these fluctuations persist for qubit 0 and qubit 4, while for qubit 2 there are additional fluctuations in the centre of the 2-state fluctuations observed at the short timescale, which lead to $\delta f$ jumping between 4 values. Finally, at the long timescales of hours (bottom row), the results exhibit a multitude of different kinds of fluctuations that can no longer be described by discrete jumps between a few states, but rather require fluctuations spanning a continuous range.}
    \label{fig:freq_diff_timescales}
\end{figure*}    
Since the main result of this article is the automatic tracking and disentanglement of time-varying $\delta f$ fluctuations, here we only give a short analysis of the results for $\Gamma_1$ and $\Gamma_\phi$. In~\figref{fig:noise_model_short_time}(b) variations in the $\Gamma_1$ at the $0.1$s timescale can be seen, although rather than discrete jumps these appear continuous. However, the 95\% confidence interval in the estimates is of similar size to that of the fluctuations, indicating that they may be due to statistical variations caused by the small number of data points in the time-dependent averaging process. To analyse whether a constant $\Gamma_1$ can agree with the observed data, the uncertainty-weighted average of $\Gamma_1$ is evaluated and depicted as the horizontal black line. One can observe that this mean is within the 95\% confidence interval of the data. The results are therefore consistent with the true value of $\Gamma_1$ being constant during the timescale shown, with the apparent variations of $\Gamma_1$ over the one second time scale due to the small sampling of measurement data and the corresponding uncertainty in the results. To further confirm this, we perform an emulation with a constant $\Gamma_1$ in the noise model and evaluate whether analogous $\Gamma_1$ fluctuations appear for analogous numbers of samples. The results are presented in~\appref{app:sec:noise_model_evaluation_simulated_data}, and show that indeed analogous fluctuations in $\Gamma_1$ are visible also in these emulations.

Although the $\Gamma_1$ fluctuations cannot be distinguished from statistical fluctuations at the sub-second timescale, we can still analyse fluctuations at longer timescales. To do this, we evaluate the power spectrum of the $\Gamma_1$ fluctuations and find that at longer timescales of 10-1000~s, they are described by $1/f^\alpha$ noise with $\alpha \sim 0.8-1.2$ (see~\appref{app:sec:gamma_1_fluctuations_analysis}). This is in agreement with the previously observed behaviour evaluated at slower time-scales~\cite{burnett2019decoherence,Klimov_2018,paladino_1f_2014,schlor2019correlating}, where fluctuations have been observed to obey a $1/f^\alpha$ power spectrum with $\alpha \sim 1$. 

In the case of $\Gamma_\phi$ (\figref{fig:noise_model_short_time}(c)), in addition to the fluctuations analogous to those observed for $\Gamma_1$ and caused by the low number of repetitions, five sharp peaks can be seen. These large jumps occur at the times at which discrete jumps in $\delta f$ occur, as illustrated by the vertical, dashed red lines. These are also seen in the simulated data presented in~\appref{app:sec:noise_model_evaluation_simulated_data}. These peaks can be explained by the fact that when averaging over a time period where a jump in frequency occurs, the resulting qubit state dynamics are described by two-frequency oscillations in $\expval{\sx}, \expval{\sy}$. A single-frequency Markovian noise model, such as the one used here, cannot generate these dynamics~\cite{agarwal2023modelling}, and instead approximates the best fit to the data by generating the peaks in $\Gamma_\phi$. To correctly describe the $\Gamma_\phi$ one can use a two-frequency model for such regimes~\cite{agarwal2023modelling}. 
Generally, when averaging over long timescales, one can expect to have many such jumps in the qubit frequency within the averaging time, as well as jumps due to fluctuations at slower timescales. In this case, the effective noise on the qubit is no longer Markovian and thus Ramsey interferometry based methods of estimating the $T_{2}$ time would fail to accurately describe the dephasing rate, since the observed decay would not be a single-exponential decay. Performing Hahn-echo based experiments, in which an echo pulse in the middle of the circuit cancels the effect of frequency fluctuations occurring at timescales longer than individual circuit durations, is typically used to mitigate the effect of such fluctuations~\cite{Hahn1950SpinEchoes}. However, it does not perform a characterisation of the fluctuations. Thus, the fast-noise-tracking methods presented here can fill this gap by characterising the fluctuations, as well as the fluctuation induced non-Markovian noise on the qubits present when averaging over long time-scales. We note that $\Gamma_1$ estimates are not affected by the frequency jumps, so significant changes in behaviour of $\Gamma_1$ around the location of frequency jumps are not expected and not seen in~\figref{fig:noise_model_short_time}(b). 

In summary, we have shown how using the few-repetition averaging protocol allows one to detect frequency fluctuations at fast timescales of tens to hundreds of milliseconds, which cause non-Markovian noise in the qubit dynamics when measurements are averaged over longer time-scales. Although one also sees fluctuations in the $\Gamma_1,\Gamma_\phi$ parameters at the fast timescales, these are largely due to the statistical random variations in measurements due to averaging over a low number of repetitions. This shows the importance of careful uncertainty quantification integrated with the non-linear regression machine learning approaches used to obtain parameters when characterising fluctuations at fast timescales. For more precise tracking of $\Gamma_1,\Gamma_\phi$ fluctuations, the averaging window size $W_G$ can be optimized for tracking those specific fluctuations instead of being optimised for tracking $\delta f$ as done in this work. The remainder of this manuscript focuses on the application of the HDFA algorithm on the time-varying fluctuations occurring in $\delta f$ and on the analysis of the obtained results.

\subsection{Frequency fluctuations at different timescales}
\label{subsec:results:freqeuency_fluctuations_diff_timescales}
The results in the previous section show that over short sub-second time-scales there is a single RTN fluctuation in the qubit frequency. The question then arises whether such behaviour persists over longer time-scales, or if there are further time-dependent changes of this behaviour. To answer this, we analyse fluctuations of $\delta f(t)$ over different timescales. In~\figref{fig:freq_diff_timescales}, the obtained values of $\delta f(t)$ are shown for the three qubits at multiple timescales. At the shortest timescales in row (a), the dominant visible fluctuations of the qubit frequencies are between two discrete values, corresponding to an RTN, occurring at timescales of a few hundred milliseconds. As also found in the previous section, a greater uncertainty in the fitted frequency value at times where there is a jump in frequency can be observed.
The standard errors for the $\delta f$ estimates for qubit 4 are significantly larger than for the other qubits. This is due to qubit 4 having a significantly larger mean $\Gamma_1$, and equivalently a much shorter $T_1 = \Gamma_1^{-1}$, than the other qubits ($\Gamma_1^{-1}$ = $115,104,21~\mu\mathrm{s}$ for the three qubits respectively), which leads to reduced resolvability of the oscillations that are fit in the non-linear regression to extract $\delta f$.

In row (b) of~\figref{fig:freq_diff_timescales}, we present results obtained over a longer timescale of $100~\mathrm{s}$. In qubit 2, transitions between four different states rather than two can be observed, occurring at slower timescales of a few seconds. These slower fluctuations shift the centre of the fast fluctuations observed at the shorter timescales. A change in the amplitude of the fast fluctuations is visible in qubit 4. These features, namely a fluctuating centre of the fast fluctuation and jumps in the magnitude of the fast fluctuations are both more clearly evident in row (c), which covers about one hour of data acquisition. The ability to detect fast fluctuations using the few-repetition averaging protocol allows analysing the stability of the qubit frequency over timescales ranging from $\sim 10^{-2}~\mathrm{s}$ to $\sim 10^{3}~\mathrm{s}$, which shows the presence of multiple concurrent fluctuations.

\begin{figure*}[tbh]
    \centering
    \includegraphics[width=\textwidth]{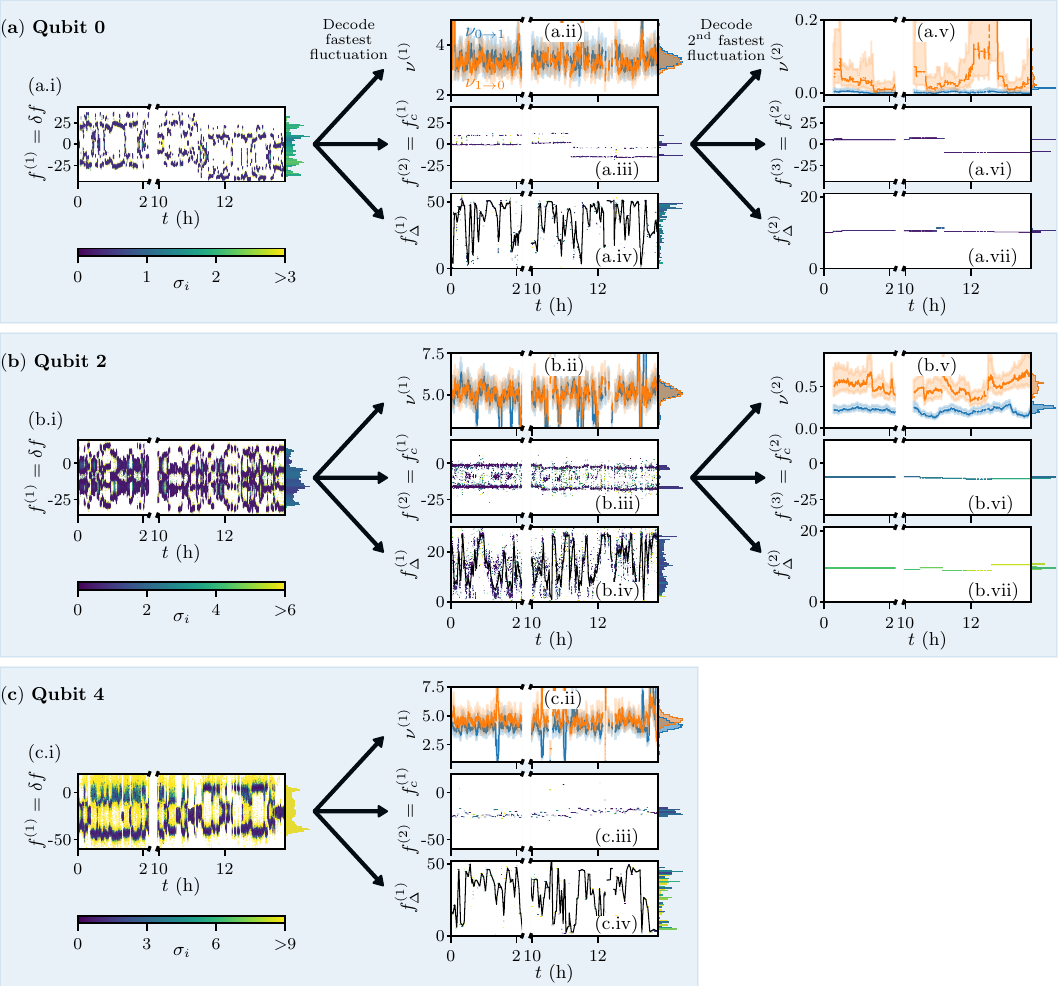}
    \caption{Results of the HDFA algorithm for the three qubits, where the columns of plots represent the increasing levels of HDFA hierarchy. Plot i for each qubit shows the $n=1$ level fluctuation $f^{(1)}$, which corresponds to the input $\delta f$, and plots iii and vi show the $f^{(2)}$ and $f^{(3)}$ HDFA fluctuations hierarchies, respectively. While the input $\delta f$ data exhibits a time-dependence that appears rather random, the underlying fluctuations get progressively disentangled at increasing levels of hierarchy, resulting in clearly visible structures in the frequency fluctuations over time in $f^{(2)}$ and $f^{(3)}$ for all qubits. For qubit 4 HDFA terminates at $n=2$ since there is no discernable RTN beyond that level. The other plots show the time-dependent rates (ii and v) and magnitudes (iv and vii) of the fluctuations disentangled at the corresponding level of hierarchy. For the fastest decoded fluctuations (middle column), the rates $\nu^{(1)}_{0\rightarrow1}$ and $\nu^{(1)}_{1\rightarrow0}$ are approximately equal, while the rates for the second fastest decoded fluctuations (third column) are asymmetric and change significantly with time. The magnitudes of the fastest fluctuations exhibit significant variation over time spanning a continuous range, while for the second fastest fluctuations the magnitudes are approximately constant. In all plots, the histogram resulting from the data in the plot is shown on the right of each plot. In the plots and histograms for the frequencies and magnitudes, the colour represents the uncertainty in the estimated parameter, and the units are in $\kHz$; in the plots for the rates the switching rates $\nu^{(n)}_{0\rightarrow1}$ and $\nu^{(n)}_{1\rightarrow0}$ and their $95\%$ confidence interval is presented in units of $\mathrm{s^{-1}}$. A rolling window average trace is shown for the $f^{(1)}_\Delta$ results in iv to guide the eye, with a window of $500~\mathrm{s}$.}
    \label{fig:disambiguation_q024}
\end{figure*}    

\subsection{Disentangling different types of fluctuations with HDFA}
\label{subsec:results_auto-segmentation}
Having observed multiple concurrent and overlapping fluctuations in the qubit frequency, the HDFA algorithm is used to disentangle the different fluctuations. The HDFA algorithm is applied to to the results for $\delta f (t)$, where we set the first level of fluctuations to $f^{(1)}(t) = \delta f (t)$. For clarity, magnitudes of fluctuations are presented in units of $\mathrm{kHz}$, while rates of fluctuations are presented in units of $\mathrm{s}^{-1}$. The full results of the HDFA algorithm used to disentangle all fluctuations is presented in \figref{fig:disambiguation_q024}. For the considered time range, the algorithm finds two levels of the fluctuation hierarchy for qubits 0 and 2, and a single level for qubit 4. 

At the first level in the fluctuation hierarchy, $\mathcal{S}^{(1)}$,the parameters $s^{(1)}(t)$, $f_c^{(1)}(t)$, and $f_\Delta^{(1)}(t)$ are evaluated from $f^{(1)}(t)$. To streamline notation, the explicit time dependence is omitted so that $s^{(1)}(t)$ is denoted as $s^{(1)}$ and so on. From the evaluated $s^{(1)}$, the switching rates of the fastest RTN is computed using a running average of $200~\mathrm{s}$ and shown in~\figref{fig:disambiguation_q024}a.ii, b.ii, and c.ii. In these plots, the rates for the $1\rightarrow 0$ and ${0\rightarrow 1}$ transitions are approximately equal and show high correlation. We verified that the large sudden dips are artefacts occurring when the fluctuation magnitude $f_\Delta^{(1)}$ is very small and hence accurately identifying $s^{(1)}$ is not possible. The mean RTN switching rates are found to be $\tilde{\nu}^{(1)}_{0\rightarrow 1} = 3.48\pm0.02, 4.68\pm0.03,3.97\pm0.03~\mathrm{s}^{-1}$ and $\tilde{\nu}^{(1)}_{1\rightarrow 0} = 3.40\pm0.02, 5.12\pm0.03,4.67\pm0.04~\mathrm{s}^{-1}$ for the three qubits, respectively. These rates confirm that the RTN switching at this fastest fluctuation level is approximately symmetric for $1\rightarrow 0$ and ${0\rightarrow 1}$ transitions ($\nu^{(1)}_{1\rightarrow 0}(t)\approx\nu^{(1)}_{1\rightarrow 0}(t)$), and occurs at similar rates for all three qubits.

We next analyse the results of the fluctuation magnitudes $f_\Delta^{(1)}$ shown in~\figref{fig:disambiguation_q024}a.iv, b.iv, and c.iv. For all three qubits, $f^{(1)}_\Delta$ shows dynamics consisting of sudden jumps as well as slow drifts occurring on timescales of a few minutes. The distributions of $f^{(1)}_\Delta$ have a constant maximum at $f^{(1)}_{\Delta \mathrm{max}} = 50.6\pm1.1,26.8\pm1.2,45.3\pm3.6\kHz$. This implies that the physical process governing the fluctuations at the first level of the hierarchy constrains the maximum magnitude of the fluctuations.

The time series of $f_c^{(1)}$ presented in~\figref{fig:disambiguation_q024}a.iii, b.iii, and c.iii themselves show discrete jumps with time. For qubits 0 and 2, a persistent RTN-like jumps are observed along with some jump(s) in the centre of these RTNs. For qubit 4 despite the large uncertainties in $f^{(1)}$, $f_c^{(1)}$ is obtained with relatively low uncertainties. Aside from the single small jump in qubit 4 $f_c^{(1)}$, no persistent fluctuations can be observed and the hierarchical model is thus terminated with all consistent fluctuations in $f^{(1)}(t)$ having been characterised. Note that longer experiments can reveal whether or not the jumps observed in~\figref{fig:disambiguation_q024}a.iii and c.iii. at $\sim 11~\mathrm{h}$ are due to a persistent and slower RTN.

For qubit 2, in~\figref{fig:disambiguation_q024}b.iii, $f_c^{(1)}$ estimates that are away from the dominant discrete states determined by the RTN are visible, although the distribution of $f_c^{(1)}$ shows that points away from the two peaks constitute only a small fraction of $<10\%$ of the data. In the distribution there is a small peak at $f_c^{(1)} \sim -10\kHz$ in the middle of two dominant peaks. These can be attributed to the cases where the RTN in $f^{(1)}$ occurs with very small magnitudes, where the true $f_\Delta^{(1)}$ is small. In these cases, the slower $f_c^{(1)}$ fluctuations are detected as the fastest fluctuations and thus are incorrectly characterised. However, as visible in the histograms, these erroneous estimates do not occur very often. Note that one can avoid these mis-classifications by using the estimates of $f_c^{(n)}$ and $f_\Delta^{(n)}$ as a prior and repeating the auto-segmentation for $f^{(n)}$ while penalising the expected incorrect peaks in the distribution of $f_c^{(n)}$. This will be subject of future work.

For qubits 0 and 2, the RTN behaviour displayed by $f_c^{(1)}(t)$ is now characterised at the next level of the HDFA fluctuation hierarchy, $\mathcal{S}^{(2)}$. For qubit 4, no persistent telegraphic noise is seen in $f_c^{(1)}(t)$ so the HDFA algorithm terminates and does not proceed to the next level. At this next level for qubits 0 and 2, the switching rates of the RTN are computed from the evaluated $s^{(2)}(t)$ using a running average of $2000~\mathrm{s}$. 
In~\figref{fig:disambiguation_q024}a.v and b.v, the switching rates fluctuate with time, although the fluctuations are of similar size to the confidence intervals.
The mean switching rates are calculated to be $\tilde{\nu}^{(2)}_{0\rightarrow 1} =0.0044\pm0.0005,0.223\pm0.006 ~\mathrm{s}^{-1}$ and $\tilde{\nu}^{(2)}_{1\rightarrow 0} =  0.028\pm0.004,0.49\pm0.02 ~\mathrm{s}^{-1}$ for qubits 0 and 2. Thus, these RTN occur at a timescale of 1-2 orders of magnitude slower than the RTN at the first level of the fluctuation hierarchy. The switching of these RTN is also significantly asymmetric for $1\rightarrow 0$ and ${0\rightarrow 1}$ transitions ($\nu^{(1)}_{1\rightarrow 0}(t)\gg\nu^{(1)}_{0\rightarrow 1}(t)$), unlike what was found for the fastest fluctuations. The extracted fluctuation amplitude of this RTN, $f_\Delta^{(2)}(t)$, remains approximately constant with time ( $f_{\Delta\mathrm{mean}}^{(2)}(t) = 10.7\pm0.2,14.7\pm0.2\kHz$ for the two qubits). The centre of the RTN $f_c^{(2)}(t)$ is also approximately constant in time except for a jump of $\sim 10\kHz$ in qubit 0.

In this section, it has been demonstrated that the developed HDFA algorithm can disentangle multiple different sources of fluctuations that affect $\delta f(t)$ at rates which span over 3 orders of magnitude, and independently characterise the properties of the disentangled fluctuations. The ability to accurately detect and disentangle even small and short lived fluctuations from a background of faster and larger fluctuations demonstrates the capabilities of our auto-segmentation algorithm.

\subsection{Physical models}
\label{subsec:results_physical_models}

The origins of frequency fluctuations can be effects such as CP switching~\cite{riste_millisecond_2013} or TLS~\cite{schlor2019correlating}. In this section we demonstrate that using the HDFA algorithm for the concurrently occurring fluctuations in the considered devices we can disentangle these two physical origins for the noise and assign the different hierarchies of fluctuations to individual physical effects.

Quasiparticle tunnelling in superconducting qubits can lead to the switching of the charge parity of the qubit between odd and even~\cite{riste_millisecond_2013}, where the two different CP states of the qubit lead to two different values of the qubit frequency, denoted here as $f_\mathrm{even}$ and $f_\mathrm{odd}$. 
The difference in qubit frequencies for the states with different charge parity depends on the environmental charge configuration and on the applied gate voltage. Together, these lead to an external voltage bias, $V_\mathrm{env}$, and a resulting charge offset, $n_g = C_gV_\mathrm{env}/2e$, where $C_g$ is the effective capacitance and the factor of $2e$ is due to the dimensionless $n_g$ being in units of cooper pair charge $2e$~\cite{serniak_direct_2019,riste_millisecond_2013}.
The absolute value of the maximum difference between $f_\mathrm{even}$ and $f_\mathrm{odd}$ is denoted as the charge dispersion, $\Delta_\mathrm{CP}$. It is largely determined by design properties of the qubit such as its frequency and anharmonicity~\cite{koch_charge-insensitive_2007}, and for the qubits on the \textit{ibmq\_lima} device one can estimate $\Delta_\mathrm{CP} \sim 20-50\kHz$ (see~\appref{app:subsec:theory_cps_max_prediction}).
The frequency difference relates to $\Delta_\mathrm{CP}$ and $n_g$ as
\begin{equation}
    f_\mathrm{even} - f_\mathrm{odd} = \frac{\Delta_\mathrm{CP}}{2} \cos (2\pi n_g).
    \label{eq:charge_dispersion}
\end{equation}
In experiments, one typically can only measure the absolute difference $|f_\mathrm{even} - f_\mathrm{odd}|$ since one does not know which of the higher or lower energy CP state corresponds to the odd or even state. Given the absolute difference, only $|n_g\ \mathrm{mod}\frac{1}{4}|$ can be evaluated due to the periodicity.

We can then determine whether the fastest observed fluctuations at level $\mathcal{S}^{(1)}$ are compatible with an origin due to CP switching by comparing the value of $\Delta_\mathrm{CP}$ obtained from the experimental data using HDFA with the known values determined from the device design parameters.
Furthermore for the frequency fluctuations quantified using HDFA to be compatible with CP switching, the histogram of the values of $f_{\Delta}$, which would indicate the charge offset drift over long timescales extracted from the fast frequency switching, needs to span an approximately continuous broad range and the maximum $f_{\Delta}$ needs to be consistent with the value of $\Delta_\mathrm{CP}$ determined by the qubit design parameters. We find that the $f^{(1)}_{\Delta}$ histogram does indeed span an effectively continuous broad range for all qubits (Fig.~\ref{fig:disambiguation_q024}). The HDFA estimate for $\Delta_\mathrm{CP}$ corresponds to the maximum observed value of $f^{(1)}_{\Delta}$, denoted as $f^{(1)}_{\Delta\mathrm{max}}$.
In Tab.~\ref{tab:cps_comparison} the theoretical estimates of $\Delta_\mathrm{CP}$ obtained from analytical approximations of the charge dispersion~\cite{koch_charge-insensitive_2007} and numerical simulations of the transmon Hamiltonian are compared with the HDFA results for all three qubits. The estimated values approximately agree with both the numerical and analytic approximations. The data therefore supports CP fluctuations as the origin.

The switching rates of CP fluctuations are not determined solely by design parameters, but also depend on the detailed nature of the qubit structure, materials, and interactions with the environment. This results in a wide range of CP switching rates reported in literature going from $\sim 0.1\mathrm{s}^{-1}$ to $1000~\mathrm{s}^{-1}$~\cite{riste_millisecond_2013,christensen_anomalous_2019,serniak_direct_2019,serniak_hot_2018,liu_observation_2024,tennant_low-frequency_2022,gordon_environmental_2022,erlandsson_2023_parity}. Our determined switching rates of about $3-5~\mathrm{s}^{-1}$ (see~\secref{subsec:results_auto-segmentation}) are within this range, further consolidating that the observed fastest fluctuations are due to CP switching. The observed symmetry between the switching rates $\nu^{(1)}_{0\rightarrow1}$ and $\nu^{(1)}_{1\rightarrow0}$ is also in agreement with symmetric switching rates for CP switching observed in literature~\cite{riste_millisecond_2013, serniak_direct_2019,liu_observation_2024}. Finally, we note that if the fluctuations at this level would not be caused by CP switching, then they would have to be caused by a source of fluctuations that also leads to a continuous distribution in the RTN magnitude with a maximum that is coincidentally similar to the charge dispersions of the qubits. Altogether, we therefore conclude that there is strong indication that the fastest fluctuations are due to CP switching. This also provides corroborating evidence to the assumption used in the past of CP switching for similar qubits in literature~\cite{shirizly_dissipative_2023}.
\begin{table}
    \centering
    \begin{tabular}{m{8em} c c c}
        \toprule
          & Qubit 0 &Qubit 2&Qubit 4\\
         \midrule 
         $\Delta_\mathrm{CP}^\mathrm{numerical}$ (kHz) & $39.8$  & $21.5$ & $32.6$ \\
         $\Delta_\mathrm{CP}^\mathrm{analytical}$ (kHz) & $48.7$  & $26.0$ & $39.7$ \\
         $f_{\Delta\mathrm{max}}$ (kHz) & $50.6\pm1.1$  & $26.8\pm1.2$ & $45.3\pm3.6$ \\
         \bottomrule
    \end{tabular}     
    \caption{Maximum RTN switching magnitude predicted from analytical approximation and numerical simulation of the transmon Hamiltonian, compared with the maximum switching magnitude obtained from HDFA. }
    \label{tab:cps_comparison}
\end{table}        

Next, potential sources of origin of the RTN at the second level of the hierarchy, $\mathcal{S}^{(2)}$, are determined. Along with CP switching, coupling to TLS is a major source of telegraphic noise in superconducting qubits~\cite{schlor2019correlating}. TLS with energy splitting close to the thermal temperature undergo random, thermally induced switching between their ground and excited states. Such TLS can either couple to the qubit directly~\cite{liu_observation_2024} or indirectly, for example by interacting with another TLS which is coupled to the qubit~\cite{schlor2019correlating}. The physical nature of the TLS can vary, and includes effects such as tunnelling atoms, tunnelling electrons, and spin or magnetic impurities~\cite{muller2019towards}. These interactions between the qubit and the TLS can be mediated by charge fluctuations, critical-current fluctuations, or magnetic field fluctuations. 

A minimal model of a qubit-TLS interaction that leads to RTN-like switching of the qubit and that has been seen to affect the qubit frequency alongside CP switching is an off-resonant charge dipole residing in the Josephson Junction barrier~\cite{martinis2005decoherence,abdurakhimov2022identification,lisenfeld_electric_2019,cole2010quantitative}, which directly interacts with the qubit~\cite{liu_observation_2024}. Such charge dipoles originate from individual electrons, atoms or groups of atoms tunnelling between local minima~\cite{muller2019towards}, or more generally mechanisms involving redistribution of charges. In what follows we assess the applicability of this model to the observed fluctuations.
With this model, thermally activated switching between the ground and excited state of the TLS leads to a shift in the qubit frequency concurrently with a jump in the qubit charge offset $n_g$. The shift in qubit frequency is due to the hybridized joint states of the qubit-TLS system, while the shift in $n_g$ is due a change in the environment charge configuration when the TLS switches state. Both these jumps depend on the properties of the qubit, the TLS, and the qubit-TLS coupling. The properties of the TLS include its asymmetry energy $\epsilon_\mathrm{TLS}$ and its tunnelling energy $\Delta_\mathrm{TLS}$, with the TLS energy $hf_\mathrm{TLS} = \sqrt{\epsilon_\mathrm{TLS}^2  + \Delta_\mathrm{TLS}^2 }$. The coupling between the qubit and TLS is governed by the component of the TLS electric dipole moment parallel to the junction electric field orientation, denoted as $d_\parallel$. For details on the model, as well as the equations relating the charge offset jump size $|\delta n_g|$ and the frequency shift of $f_\Delta^{(2)}$ to the model parameters, see~\appref{app:subsec:charge_dipole_model_derivation}.

To evaluate whether this model is consistent with the observed fluctuations at the level $\mathcal{S}^{(2)}$ we determine whether there is a measurable change in $n_g$ where there is a frequency jump at level $\mathcal{S}^{(2)}$. In Ref.~\cite{liu_observation_2024} the $n_g(t)$ for the different TLS states was extracted by computing the distribution of the qubit frequencies obtained from qubit spectroscopy experiments, relying on the assumption that TLS state changes significantly faster than the charge offset variations. By virtue of the fast-tracking and HDFA algorithm, we do not rely on this assumption and instead use the values of $n_g$ obtained using~\eqref{eq:charge_dispersion} just before and after a discrete jump at level $\mathcal{S}^{(2)}$ to calculate the change in $n_g$ upon a change of the TLS state. In \appref{app:subsec:charge_offset_extraction} it is shown that there is indeed a measurable change in $n_g$ of $|\delta n_g| = 0.0014\pm0.0004, 0.011\pm0.002$ for qubits 0 and 2, respectively. This correlation between charge offset jumps and the RTN fluctuations indicates that the fluctuations at level $\mathcal{S}^{(2)}$ are due to coupling to a charge-dipole TLS that induces charge fluctuations on the qubit, and that other microscopic mechanisms of the qubit-TLS coupling such as critical current or magnetic field fluctuations can be ruled out. We note that the qubit off-resonantly coupled to a charge-dipole TLS is a minimal model that allows us to explain the observed concurrent $n_g$ and qubit frequency fluctuations; however, the qubit coupling to a charge-dipole TLS defect can also have other more indirect forms, such as the qubit interacting with a resonant TLS which itself couples to a charge-dipole TLS. For a discussion comparing direct coupling to a TLS with indirect coupling mediated by a near-resonant TLS, see Ref.~\cite{liu_observation_2024}.

To further confirm applicability of the off-resonantly coupled charge dipole model, the physical parameters of the TLS in the model are computed to determine whether these lie in ranges expected for such systems. To fully evaluate the model, the TLS energy $f_\mathrm{TLS}$ and junction thickness $x$ would need to be known. Approximate values for the TLS energy can be obtained using the ratio $\nu_{1\rightarrow0}^{(2)}/\nu_{0\rightarrow1}^{(2)}$ by assuming that the TLS is in thermal equilibrium with the environment. Although the temperature of the TLS environment, $T_\mathrm{TLS}$ is not known, it must be larger than the refrigerator temperature. Assuming the TLS environment temperature is below $100~\mathrm{mK}$ based on observations in literature~\cite{lucas2023}, we arrive at the range of parameters shown in Table~\ref{tab:extracted_properties_tls}. Using these values, along with the typical values of $x$ observed in literature, all model parameters are obtained and listed in Table~\ref{tab:extracted_properties_tls} (see~\appref{app:subsec:charge_dipole_model_derivation} for details). 

We now analyse whether the computed TLS parameters are compatible with their ranges reported in literature. In amorphous alumina, TLS defects have been found to have a parallel projected dipole moment distribution of mean $0.96e\text{Å}$ and standard deviation $0.52e\text{Å}$~\cite{hung2022probing}. In qubits designed and chosen for optimal performance there is an inherent bias towards qubits which do not have strongly coupled defects due to detrimental effects on performance of such coupled defects. Thus, it is expected that the defects observed in such qubits have weaker coupling, corresponding to lower parallel dipole moments than the typical dipole moments observed in devices without such a bias. For the charge-dipole TLS model, the estimated dipole moment components presented in Table~\ref{tab:extracted_properties_tls} are between $0.04$ and $0.45(e\text{Å})$. These fall within the distribution reported in Ref.~\cite{hung2022probing} and are less than the typical parallel dipole moments, as expected due to the bias towards weakly coupled TLS in qubits optimized for applications. The asymmetry and tunnelling energies of the TLS in this model are also in approximate agreement with the energies found in prior literature~\cite{muller2019towards}. The off-resonantly coupled charge TLS model therefore agrees well with the observed data. 

In summary, we conclude that frequency fluctuations observed in the qubit are likely due to concurrent CP switching and switching of a charge-dipole TLS, with the fast fluctuations due to CP switching and the slower fluctuations due to TLS. The off-resonantly coupled charge-dipole TLS defect model is consistent with the data. The quantitative mapping of the observed fluctuations to the physical models is made possible by the fast-fluctuation-tracking and HDFA fluctuation disentangling tools developed in this work, since it requires detecting small and short fluctuations as well as carefully disentangling such fluctuations from those occurring at other time-scales.
\begin{table}
    \centering
    \begin{tabular}{m{7em} c c}
        \toprule
          & Qubit 0 &Qubit 2\\
         \midrule 
         $|\delta n_g| (2e)$ & $0.0014\pm0.0004$  & $0.011\pm0.002$ \\
         $f_\mathrm{TLS}$ (GHz) & $0.38-3.8$  & $0.16-1.6$ \\
         $\Delta_\mathrm{TLS}/h$ (GHz) & $0.36-3.0$  & $0.07-0.56$ \\
         $\epsilon_\mathrm{TLS}/h$ (GHz) & $0.13-2.4$  & $0.15-1.5$ \\
         $d_\parallel (e\text{Å})$ & $0.04-0.16$  & $0.22-0.45$ \\
        \bottomrule
    \end{tabular}     
    \caption{Charge dipole TLS model parameters evaluated for the observed HDFA results. The ranges presented correspond to assumptions on the junction thickness being $x = 1- 2~\mathrm{nm}$ and the TLS environment temperature being $T_\mathrm{TLS} = 10-100~\mathrm{mK}$.}
    \label{tab:extracted_properties_tls}
\end{table}        

\section{Discussion}
\label{sec:discussion}
We present the HDFA framework for the fast-tracking and disentangling of different sources of noise fluctuations in qubits and use it to characterise noise in a superconducting qubits device. The few-repetition Gaussian averaging protocol allows detecting frequency fluctuations in the qubits with temporal resolution of 10s of milliseconds and tracking them automatically over many hours. We observe a multitude of different fluctuations in the qubit frequency and use HDFA to disentangle the different sources of concurrently occurring fluctuations. This enables individually characterising the different fluctuations as well as correlating the different fluctuations, which in turn allows characterisation of the noise sources through analysing the time dependence of fluctuations. For the considered transmon qubits these are mapped to compatible physical models consisting of CP switching as well as coupling to charge-dipole TLS.
Thus, the developed framework addresses the need of characterising qubit noise fluctuations caused by multiple sources with high temporal resolution, as required for optimisation of qubit operations for successfully executing quantum algorithms on near term noisy devices, as well as for achieving fault tolerant operation using quantum error correction.

The framework is also applicable to other types of fluctuations and platforms where similar fluctuations to the ones occurring in superconducting qubits may occur.
Other types of fluctuations can include those in higher qubit levels induced by CP switching: even in cases where CP fluctuations at the ground state and first excited state level are mitigated through device design, greater impact of CP fluctuations on the higher levels usually persists~\cite{koch_charge-insensitive_2007, kehrer_improving_2024} and can have significant detrimental effect on qubit performance~\cite{papic_charge-parity_2024}.
Other types of platforms include silicon qubits~\cite{peters1999random}, as well as quantum sensors where the measurement of system fluctuations is the core to the sensing functionality~\cite{echternach2018single,dixit2021searching}.
An avenue for future work involves utilising the disentangled fluctuations to mitigate the impact of fluctuations by updating the qubit calibration on the fly in real-time. This complements recent work on TLS noise mitigation in which the effects of fluctuating resonant TLS are mitigated by performing real-time TLS control~\cite{chen2025scalablesitespecificfrequencytuning,dane2025performancestabilizationhighcoherencesuperconducting}.

\section{Acknowledgements}
We thank Valerie Livina for useful discussions. All authors acknowledge the support of the UK government Department for Science, Innovation and Technology (DSIT) through the UK National Quantum Technologies Programme. We acknowledge the Horizon Europe European Metrology Partnership project 23FUN08 MetSuperQ where NPL is supported by UKRI grant number 10133632. A.A. acknowledges support through an Investigator Award from the National Physical Laboratory’s Directors’ Science and Engineering Fund. We acknowledge the use of IBM Quantum services for this work.
\input{appendix.tex}

\bibliography{references}

\end{document}

%% file: appendix.tex
\appendix
\section{Experiment and device details}
\label{app:sec:experiment_device_details}
\subsection{Device details}
\label{app:subsec:device_details}
\begin{figure}[!ht]
    \centering
    \resizebox{0.2\textwidth}{!}{%
    \begin{circuitikz}
    \tikzstyle{every node}=[font=\LARGE]
    \node [font=\LARGE] at (3,10) {0};
    \node [font=\LARGE] at (5,10) {1};
    \node [font=\LARGE] at (7,10) {2};
    \node [font=\LARGE] at (5,8) {3};
    \node [font=\LARGE] at (5,6) {4};
    \draw [line width=0.5mm,red] (3,10) circle (0.75cm);
    \draw  (5,10) circle (0.75cm);
    \draw [line width=0.5mm,red]  (7,10) circle (0.75cm);
    \draw  (5,8) circle (0.75cm);
    \draw [line width=0.5mm,red]  (5,6) circle (0.75cm);
    \draw (3.75,10) to[short] (4.25,10);
    \draw (5.75,10) to[short] (6.25,10);
    \draw (5,9.25) to[short] (5,8.75);
    \draw (5,7.25) to[short] (5,6.75);
    \end{circuitikz}
    }%
    \caption{Qubit layout of the \textit{ibmq\_lima} device. The different nodes represent the different physical qubits, and connections between the nodes represent the presence of physical coupling between the qubits which enables two-qubit gate operations. In the experiments, identical circuits were run in parallel on non-neighbouring qubits 0,2, and 4 which are highlighted in red.}
    \label{fig:ibmq_lima_topology}
\end{figure}
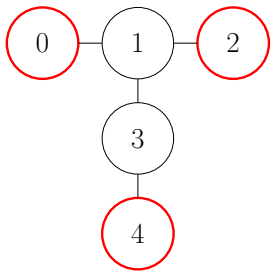

The experiments are run on the \textit{ibmq\_lima} device accessed through the IBM Quantum platform~\cite{ibm_quantum}. The device consists of 5 qubits, with the topology as shown in \figref{fig:ibmq_lima_topology}. Identical circuits are run in parallel on non-neighbouring qubits 0, 2, and 4, which minimizes the effects of static ZZ crosstalk~\cite{tripathi2022suppression}. The reported properties for the three qubits are shown in Table \ref{tab:qubit_properties}.

\begin{table}
    \centering
    \begin{tabular}{|c|c c c|}
         \hline
          & Qubit 0 &Qubit 2&Qubit 4\\
         \hline 
         Frequency $f_0$ (GHz) & $5.030$  & $5.247$ & $5.092$ \\
         Anharmonicity $\alpha$ (MHz) & $-0.336$  & $-0.334$ & $-0.334$ \\
         $X_{\pi/2}$ duration $(\mathrm{ns})$  & $35.56$  & $35.56$ & $35.56$\\
         Reset duration $(\mathrm{ns})$  & $6304$  & $6304$ & $6304$\\
         Readout duration $(\mathrm{ns})$  & $5913$  & $5913$ & $5913$\\
         \hline
    \end{tabular}     
    \caption{Properties and calibration data of the different qubits on the \textit{ibmq\_lima} device, obtained from the IBMQ platform at the time of the experiments.}
    \label{tab:qubit_properties}
\end{table}        

\section{Performance of the Gaussian moving window averaging procedure}
\label{app:sec:gaussian_window_averaging}

\begin{figure}
    \centering
    \includegraphics[width=\columnwidth]{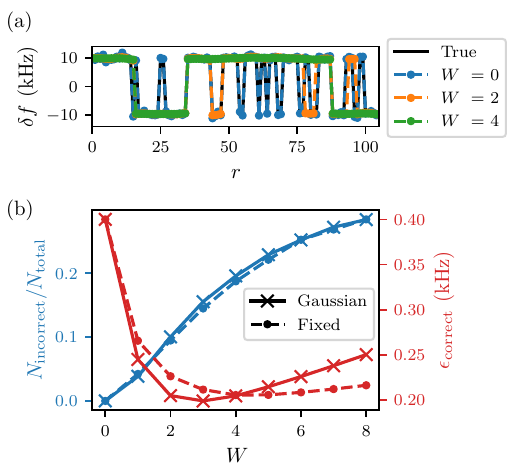}
    \caption{(a): Example of the results of $\delta f$ obtained via the Gaussian averaging procedure for simulated data for different choices of window size $W_G$. The black solid line denotes the true value of $\delta f$ while the different colours represent the estimated $\delta f$ obtained after fitting the noise model to the data obtained after averaging with different window sizes. Larger window sizes can be seen to result in missing fluctuations that happen within the window timescale. (b) Comparison of the fixed window averaging procedure with the Gaussian window averaging procedure. The variable $W$ corresponds to $W_G$ and $W_F$ in the case of the Gaussian (crosses) and fixed window (dots) averaging, respectively. The prediction inaccuracy $N_\mathrm{incorrect}/N_\mathrm{total}$ increases with $W$ for both the Gaussian and fixed averaging procedures due to the missing of fluctuations as seen in (a). The median error of correct predictions, $\epsilon_\mathrm{correct}$, reduces faster with $W$ for the Gaussian averaging procedure than for the fixed window approach.}
    \label{fig:gaussian_vs_fixed}
\end{figure}    

In order to evaluate the performance of the Gaussian window averaging procedure described in Sec. \ref{subsec:gaussian_averaging} (Eq. \eqref{eq:gaussian_moving_average_def}), in this section we compare the results obtained using Gaussian averaging  with that of the fixed window moving averaging procedure defined below and evaluate the shortest jump that can be detected by the method. 

In the fixed window moving averaging protocol, a moving window averaging with window size $2W_F + 1$ is performed, such that the measurements results in the integer $W_F$ repetitions preceding and succeeding $r$ are averaged along with the results for repetition $r$ to estimate $p(\tau,b,t_r)$. For the fixed moving average, the estimated probability is given by
\begin{equation}
    p(\tau,b,t_r) = \frac{1}{2W_F + 1} \sum_{i=r-W_F}^{r+W_F} {B(\tau,b,t_i)}. 
    \label{eq:fixed_moving_average_def}
\end{equation}    

In order to compare the two methods, here we use fixed parameters in a noise model and perform emulator runs to systematically evaluate the relative results of the two methods. We simulate a noise characterisation experiment with the qubit frequency detuning fluctuating between two values, $\delta f (r) = s(r)f_\Delta/2$, with $r$ representing the repetition index as in the main text, $s(r) \in \{ +1, -1 \}$, $s(r)$ with an average switching rate of one every $5$ repetitions, and $f_\Delta = 20\kHz$. The parameters $\Gamma_1(r), \Gamma_\phi(r)$ are both set to $0$. We perform averaging with both averaging methods for different values of $W_G$ and $W_F$, and then perform non-linear regression to estimate the noise model parameter $\delta f (r)$. Then, a single level of the HDFA algorithm is run to evaluate the most likely values of $s(r)$, denoted $\tilde{s}(r)$.

An example of results obtained for the Gaussian averaging procedure is shown in \figref{fig:gaussian_vs_fixed} (a), and compared with the known time-evolution used in the emulator runs. Larger values of $W_G$ result in more precision in estimates of $\delta f$, but they can miss fluctuations that occur inside the moving window average size. To compare with the fixed window averaging procedure, we compute metrics for both these effects. For evaluating the effect of missing fluctuations, the predicted states $\tilde{s}(r)$ are compared with the true states $s(r)$ and the number of time steps for which $\tilde{s}(r) \neq s(r)$, denoted as $N_\mathrm{incorrect}$, is computed. The ratio $N_\mathrm{incorrect}/N_\mathrm{total}$, where $N_\mathrm{total}$ is the total number of time steps in the emulation, is then used as a metric that quantifies the number of missed fluctuations. For a metric quantifying the precision of the obtained results, we evaluate the difference between the true and estimated values of $\delta f$. Note that we only do this for the subset of results for which assigned state $s(r)$ is correct, since the incorrect classifications are already accounted for in the first metric. The median difference, $\epsilon_\mathrm{correct}$, is used as the metric for the precision.

In \figref{fig:gaussian_vs_fixed} (b), we observe that the fraction of incorrect classifications increases with both $W_G$ and $W_F$, and is very similar for both windowing methods. This indicates that both methods have similar temporal resolution. For the precision, we see that $\epsilon_\mathrm{correct}$ first reduces with $W_G$ and $W_F$, as expected due to reduced statistical noise that scales as $\sim 1/\sqrt{W_{G/F}}$. However, instead of continuing to reduce, $\epsilon_\mathrm{correct}$ reaches a minimum and then starts increasing again. This is explained by longer windowing lengths having lower time resolution and being more likely to include a $\delta f$ jump in the averaging window, leading to more model violation. Thus, after $\epsilon_\mathrm{correct}$ reaches a minimum, the error is dominated not by limited statistical precision but rather by model violation due to low temporal resolution. Note that the specific location of this minimum is very dependent on the fluctuation rate, since slower fluctuations allow longer averaging length before the fluctuations dominate the error. The Gaussian procedure results in a lower imprecision $\epsilon_\mathrm{correct}$ reached for small window sizes. Thus, in the regime where the likelihood of discrete jumps within the averaging window is low, the emulation results indicate that the Gaussian averaging method outperforms the fixed averaging procedure in precision while having the same time resolution.

\begin{figure}
    \centering
    \includegraphics[width=\columnwidth]{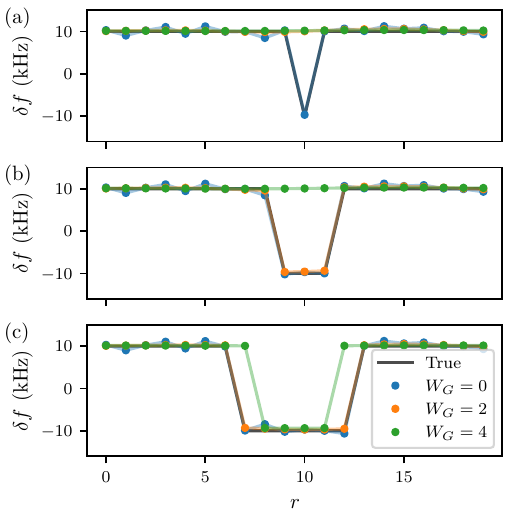}
    \caption{Simulated results for a fluctuation lasting for different durations. The solid black line represents the true value of $\delta f$ as set in the emulator runs. In the top row, we can observe that only the results for $W_G = 0$ detect the fluctuation, while the others miss the fluctuation entirely. In the middle row, we see that a fluctuation lasting for 3 repetitions is detected by the $W_G=0,2$ results but not by the $W_G=4$ results. Lastly, in the bottom row we observe that a fluctuation lasting for $6$ repetitions is detected by all three window sizes.}
    \label{fig:simulated_data_fluctuation_missing}
\end{figure}

In order to evaluate the fastest fluctuation that we can detect with the Gaussian averaging procedure, in \figref{fig:simulated_data_fluctuation_missing} we show example results where a $\delta f$ jump occurs for a duration of $1,3,$ and $6$ repetitions. For $W_G = 0$, which corresponds to averaging just a single repetition, all fluctuations are detected correctly. This is expected since for each repetition $r$, only the results for that repetition are used and thus the estimated parameter $\delta f$ is effectively evaluated independently each time. However, due to large uncertainties in the $W_G=0$ results because of the insufficient number of samples used in the averaging, we also see instances where the estimated $\delta f$ is significantly different from the true value. For $W_G=2$ one can observe that a fluctuation lasting for just 1 repetition is missed, but that a fluctuation lasting for $\geq 3$ repetitions is captured correctly. Similarly, one can observe that for $W_G = 4$, fluctuations lasting for $\leq 3$ repetitions are missed entirely. A fluctuation lasting for $6$ repetitions is indeed detected by the $W_G=4$ model. However, the estimated fluctuation starts later and ends sooner than the true fluctuation.

Thus, for the noise characterisation experiments performed in this work, with the few-repetition Gaussian averaging procedure with the window size $W_G = 2$ we can detect discrete $\delta f$ jumps that last for for at least $3$ repetitions, corresponding to tens of milliseconds time resolution in the results presented in the main text.

\section{Computing uncertainty in the noise model parameters}
\label{app:sec:frequency_uncertainty_estimation}
This section presents how the uncertainties in the estimates of the noise model parameters are computed. Bootstrap methods are widely used as a non-parametric approach to estimate standard errors in parameters~\cite{efron1994introduction}. In the context of model fitting, the general approach involves creating new samples of data from the original data, performing the model fitting for all the new samples, and evaluating the standard deviation of the distribution of the evaluated model parameters. This standard deviation is an estimate of the standard error in the estimate of the evaluated model parameter.

The methods used to create the new samples from the data determine the sources of errors that are factored in the estimate of the total uncertainty. In the context of noise model fitting with few-repetition-averaged data, the two primary sources of uncertainties that need to be taken into account are the imperfect model fitting as well as statistical imprecision due to few-repetition averaging. For the former, residual resampling, which involves creating new samples which differ from the original depending on how well the model fit the original data, can be used~\cite{hesterberg2011bootstrap}. For the latter, new samples can be created by drawing new estimates of the probabilities from the expected binomial distribution, thus accounting for the statistical error due to few-repetition sampling. In this work both of these resampling methods are combined to generate new samples which are affected by both the imperfect model fitting as well as the statistical few-repetition-averaging errors.

In each new bootstrap sample, first the statistical error is taken into account by transforming the original Gaussian-averaged probability $p(\tau,b,t_r)$ to a new probability evaluated from a binomial distribution corresponding to probability $p(\tau,b,t_r)$ and a number of trials $N_\mathrm{eff}$, where $N_\mathrm{eff}$ is the effective number of trials given by $N_\mathrm{eff} = \sum_i w(i,r)$, with $w(i,r)$ being the weight in the Gaussian averaging defined in~\eqref{eq:gaussian_averaging_weight_definition}. Thus, the transformation is given by
\begin{equation}
p(\tau,b,t_r) \rightarrow \mathcal{B}(p(\tau,b,t_r),N_\mathrm{eff})/N_\mathrm{eff}.
\end{equation}

After this transformation for each $p(\tau,b,t_r)$, the error due to the imperfect model fitting is taken into account in the resample.  Let $p_\mathrm{fit}(\tau,b,t_r)$ be the probability obtained from the noise model after performing non-linear regression corresponding to the original probability $p(\tau,b,t_r)$.
The residuals corresponding to the difference between the original data and the estimated data obtained from the model fitting are given by $\epsilon(\tau,b,t_r) = p(\tau,b,t_r) - p_\mathrm{fit}(\tau,b,t_r)$. To generate the new samples, the residuals are added to the transformed probabilities by randomly choosing from all the residuals with replacement. Let $\mathcal{R}(\epsilon(\tau,b,t_r))$ represent a randomly chosen value from the set of all residuals $\epsilon(\tau,b,t_r)$, the final probabilities in the new sample, denoted $\tilde{p}$, are given by
\begin{equation}
\tilde{p}(\tau,b,t_r) = \mathcal{B}(p(\tau,b,t_r),N_\mathrm{eff})/N_\mathrm{eff} + \mathcal{R}(\epsilon(\tau,b,t_r)).
\end{equation}

Note that adding the residual can lead to $\tilde{p}(\tau,b,t_r) <0$ or $\tilde{p}(\tau,b,t_r) > 1$ in the transformed samples. To account for this, probabilities smaller that $0$ are set to $0$ and probabilities larger than $1$ are set to $1$.

Once $100$ bootstrap samples are created, the noise model is fit to all the samples and the model parameters $\delta f, \Gamma_1, \Gamma_\phi$ are computed for each sample. The standard deviation of the distribution of $\delta f, \Gamma_1, \Gamma_\phi$ for all samples is used as the estimate of the uncertainties $\sigma_{\delta f}, \sigma_{\Gamma_1}, \sigma_{\Gamma_\phi}$ respectively.

\section{Testing noise model evaluation procedure with simulated data}
\label{app:sec:noise_model_evaluation_simulated_data}
\begin{figure*}
    \centering
    \includegraphics[width=\textwidth]{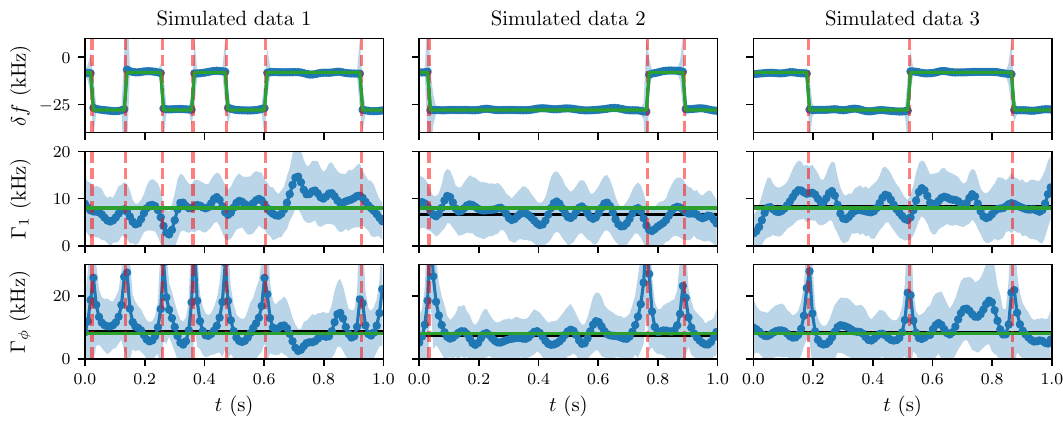}
    \caption{(a): Example of three sets of noise model evaluation results with simulated data. The green lines denote the true noise model parameters in each simulation, while the blue dots denote the evaluated noise model parameters with the shaded region denoting the $95\%$ confidence interval. The horizontal black lines in $\Gamma_1$ and $\Gamma_\phi$ denote the weighted mean of the estimated respective noise parameters. The vertical red dashed lines denote the locations of a frequency fluctuations, which can be seen to lead to a jump in the evaluated $\Gamma_\phi$ parameter.}
    \label{fig:noise_model_simulated_data}
\end{figure*}    
In this section we present results for numerical simulations of the noise characterisation experiments to facilitate comparison with experimental results. The noise model used in the simulations to generate the raw data is identical to the noise model described in~\secref{subsec:noise_model_evaluation} with the values of the noise parameters in the simulations chosen to be similar to the values obtained from the experimental results presented in~\secref{subsec:noise_model_evaluation_results}. The parameters $\Gamma_1(t_r),\Gamma_\phi(t_r)$ are set to be constant and equal to $8\kHz$, while $\delta f(t_r)$ exhibits discrete jumps between values of $+2\kHz$ and $-28\kHz$ with a switching probability of $1/20$ at every repetition. From these noise parameters, the true probabilities $p(\tau,b,t_r)$ are evaluated and then from the true probabilities, binary measurements $B(\tau,b,t_r)$ are sampled from the binomial distributions. Then, Gaussian averaging of the raw data is performed with $W_G=2$ and the noise model is fit to the averaged data for all $t_r$. The results for three sets of the simulations with different random seeds are presented in~\figref{fig:noise_model_simulated_data}. 

The results in~\figref{fig:noise_model_simulated_data} exhibit very similar features to the experimental results shown in~\figref{fig:noise_model_short_time}, including clean detection of the discrete jumps in $\delta f$, fluctuations in $\Gamma_1$ which are of similar size to the confidence intervals, and peaks in $\Gamma_\phi$ which are co-located with jumps in $\delta f$. Thus, this corroborates that the fast-tracking method successfully obtains the $\delta f$ fluctuations, while the fluctuations of $\Gamma_1$ and $\Gamma_\phi$ at the sub-second timescales observed in the experiments are due to statistical uncertainty due to the small number of repetitions, as discussed in the main text in~\secref{subsec:noise_model_evaluation_results}.

\section{\texorpdfstring{$\Gamma_1$}{Relaxation rate} power spectral density}
\label{app:sec:gamma_1_fluctuations_analysis}
\begin{figure}
    \centering
    \includegraphics[width=\columnwidth]{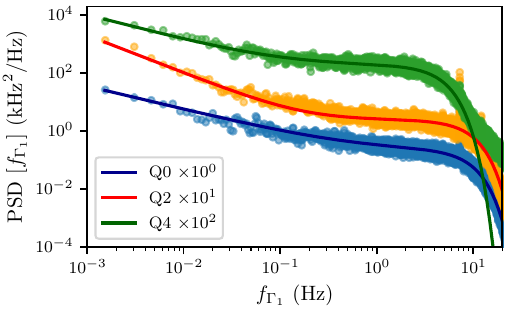}
    \caption{Power spectral density (PSD) of $\Gamma_1$ fluctuations for the three qubits. The solid lines represent fits of the PSD to a white noise and $1/f^\alpha$ model. Note that the behaviour of the PSD for $f_{\Gamma_1} \gtrsim 10~\mathrm{Hz}$ is dominated by the effects of Gaussian window averaging.}
    \label{fig:ad_power_spectrum}
\end{figure}    
This section analyses the $\Gamma_1$ fluctuations in the frequency domain. Power spectral density has been previously used to characterise the stability of qubit relaxation times~\cite{burnett2019decoherence}. Following the method outlined in~\cite{burnett2019decoherence}, this quantity, denoted as $\mathrm{PSD}[f_\mathrm{\Gamma_1}]$, is computed for all thee qubits and presented in ~\figref{fig:ad_power_spectrum}. The PSD is obtained using the Welch method~\cite{welch2003use} with a segment size of $2^{16}$.

In the PSD, the power of fluctuations reduces with increasing frequencies until about $f_{\Gamma_1} \sim 10^{-1}-10^0 \mathrm{Hz}$, at which point the PSD flattens to a constant value. Then the PSD maintains the constant value until $f_{\Gamma_1} \sim 10^1 \mathrm{Hz}$ at which point there is a sharp reduction in the PSD. The intial decrease in PSD at low frequencies is explained by a power law decay of the PSD, as observed in~\cite{burnett2019decoherence}. The proceeding flattening corresponds to white noise in the fluctuations. This is likely caused by errors in the estimation of the parameter $\Gamma_1$ as observed in the numerical simulations in~\appref{app:sec:noise_model_evaluation_simulated_data}. The final sudden reduction in the PSD is due to the Gaussian averaging of the data, which leads to autocorrelations in the estimates of $\Gamma_1$ at frequencies corresponding to the averaging window size.

To model these fluctuations, a functional form of the fluctuations with a $1/f^\alpha$ noise and white noise is fit to the PSD data. The effects of the autocorrelation introduced by the Gaussian averaging are taken into account without introducing additional fitting parameters by convolving the model PSD with the Gaussian averaging kernel. The results of the model fitting are shown in~\figref{fig:ad_power_spectrum} where the model shows very good agreement with the data. The values of $\alpha$ for the three qubits are $0.81,1.30,0.87$ for qubits 0,2, and 4, respectively. This is in agreement with the observed approximately $1/f^\alpha$ noise in the relaxation rate fluctuations~\cite{paladino_1f_2014,burnett2019decoherence}. 

\section{Hierarchical discrete fluctuation removal}
\label{app:sec:auto-segmentation}
\subsection{Automated hyperparameter evaluation of HDFA algorithm}
\label{app:subsec:automated_hyperparameter_evaluation}
\begin{figure*}
    \includegraphics[width=\textwidth]{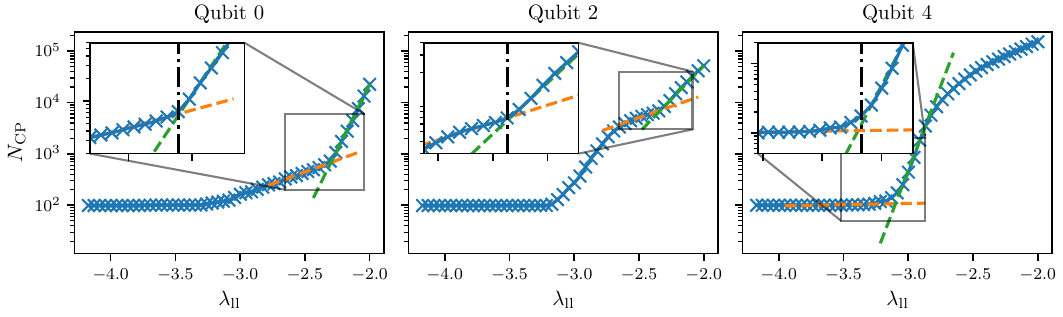}
    \caption{Elbow plots for determining the hyperparameter $\lambdall$ for the three qubits. As the threshold $\lambdall$ is increased, the quality criteria for the HMM becomes stricter and the auto-segmentation results in more segments, and thus, a higher number of change points $N_\mathrm{CP}$. The vertical dashed black lines in the insets denote the chosen value of $\lambdall$ which is where a kink in the $N_\mathrm{CP}$ can be seen.}
    \label{fig:choose_gammma_ll}
\end{figure*}
\begin{figure*}
    \includegraphics[width=\textwidth]{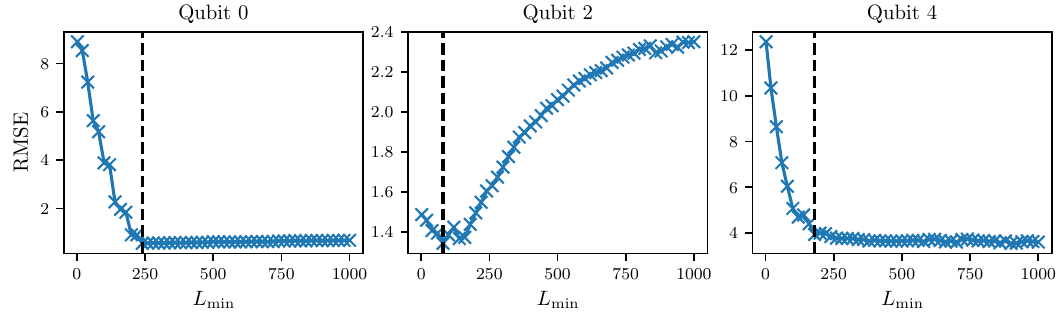}
    \caption{The accuracy of the results of the auto-segmentation as a function of the minimum segment length hyperparameter, $L_\mathrm{min}$, for the three qubits. The chosen values of $L_\mathrm{min}$ is depicted via the vertical dashed black lines.}
    \label{fig:choose_msl}
\end{figure*}
The auto-segmentation algorithm has two hyperparameters: $\lambdall$ and $L_\mathrm{min}$. In this section we present our method for evaluating appropriate choices for these parameters. The method involves choosing these parameters independently: first we evaluate an appropriate choice of $\lambdall$, then we evaluate $L_\mathrm{min}$. We denote our chosen hyperparameters as $\lambdall^{(o)}$ and $L_\mathrm{min}^{(o)}$.

In order to choose $\lambdall$, we first set $L_\mathrm{min} = 2$, allowing arbitrary small segments. As the threshold $\lambdall$ is varied, in the limit of $\lambdall\rightarrow 0$ (its maximum value), the quality threshold that HMMs need to meet for a new data point to be added to a segment becomes stricter. Thus, in the limit of large $\lambdall$, the number of segments resulting from the auto-segmentation algorithm, denoted $N_\mathrm{CP}$, increases. In the opposite limit, the data is divided into only a few segments. If the threshold fitting quality is stricter than the typical fitting quality for ``true'' segments, the true segment is broken into multiple ones, resulting in a rapid increase in $N_\mathrm{CP}$ with $\lambdall$. The location of this rapid increase is visible as an elbow in the plot of $N_\mathrm{CP}$ against $\lambdall$ and we choose this location as $\lambdall^{(o)}$. For the data in this manuscript, the plots depicting this choice for the first level of the HDFA algorithm are shown in~\figref{fig:choose_gammma_ll}.

Once we have chosen $\lambdall$, we need to choose $L_\mathrm{min}$. Ideally, $L_\mathrm{min} = 2$ should be chosen so that arbitrarily small segments are allowed. However, if $L_\mathrm{min}$ is too small, HMM segments can be terminated too early when not sufficient statistics on the fastest RTN are included in the segment. If $L_\mathrm{min}$ is too large, true segments smaller than $L_\mathrm{min}$ are not segmented correctly. To find the optimal value of $L_\mathrm{min}$, we perform the auto-segmentation with the fixed value of $\lambdall = \lambdall^{(o)}$ for a range of $L_\mathrm{min}$ values. As a quality metric, we use the root-mean-squared-error (RMSE) between the original data $f^{(n)}(t)$ and the data predicted after the auto-segmentation $\tilde{f}^{(n)}(t)$. We then use the minimum value of $L_\mathrm{min}$ that leads to the lowest RMSE, or the value at which the RMSE stops changing significantly, as the value of $L_\mathrm{min}^{(o)}$. For the data in this manuscript, the plots depicting this choice for the first level of the HDFA algorithm are shown in~\figref{fig:choose_msl}.

\subsection{Computing properties from HDFA results}
\label{app:subsec:hdfa_computing_properties}

\subsubsection{Extracting uncertainties in HDFA results}
\label{app:subsec:auto-segmentation_uncertainty}

In order to estimate the total uncertainties in parameters $f^{(n)}_c$ and $f^{(n)}_\Delta$ resulting from the auto-segmentation algorithm, we consider two different sources of errors. The first error corresponds to the fitting quality of each segment resulting from the segmentation while the second takes into account different potential segmentations of the data. For ease of notation and clarity, we neglect the superscript ``${(n)}$" in the discussion below.

Let $\mathcal{T}$ correspond to a single segment predicted by the auto-segmentation algorithm. Then, $f(t_r\in \mathcal{T})$ denotes the ``true'' data. Let $s^{\mathcal{T}}$, $f^{\mathcal{T}}_c$ and $f^{\mathcal{T}}_\Delta$ denote the predicted RTN state, centre value and fluctuation size, respectively, for this segment. For the fitted HMM, the predicted values for $f(t_r)$ are $\tilde{f}(t_r) = f^{\mathcal{T}}_c + s^\mathcal{T}(t_r)f^{\mathcal{T}}_\Delta/2$. We also have uncertainties in the estimates of $f(t_r)$, denoted $\sigma_{f(t_r)}$, which need to be appropriately propagated into the estimate of the total uncertainty. By virtue of the HMM, each segment is itself divided into periods of constant $\tilde{f}(t_r)$ when consecutive $s^\mathcal{T}(t_r)$ are identical. Let the number of times that $s^\mathcal{T}(t_r)$ changes value be $M-1$, so that the segment can effectively be divided into $M$ blocks corresponding to periods of constant $\tilde{f}(t_r)$. Let $\mathcal{T}_{m=1,\ldots,M}$ denote the times corresponding to these blocks.

For each block $m$, we define $f_m$ and $\sigma_{f_m}$ be the weighted mean of $f(t_r\in \mathcal{T}_m)$ and its standard error, respectively, where the weights in the mean are related to the uncertainties in the values as $1/\sigma^2_{f(t_r)}$. This weighted mean and the standard error in the weighted mean propagates the uncertainty in the estimates of $f(t_r\in \mathcal{T}_m)$. The variability of $f_m$, along with the uncertainties in each estimated $f_m$ need to both be accounted for in order to obtain an estimate of the uncertainty in $f_c^\mathcal{T}$ and $f_\Delta^\mathcal{T}$. This is done by performing a weighted standard deviation of the differences and means of consecutive values of $f_m$, with the weights determined by $\sigma_{f_m}$. The measures of the uncertainties in $f_c^\mathcal{T}$ and $f_\Delta^\mathcal{T}$ are then given by
\begin{equation}
\begin{split}
\sigma^{(1)}_{f_c}(t \in \mathcal{T}) = \mathrm{WSD}\Biggl(&\Bigl\{\frac{f_m + f_{m+1}}{2} \Bigr\}_{m=1}^{M-1}, \\&\Bigl\{\frac{\sigma_{f_m} + \sigma_{f_{m+1}}}{2} \Bigr\}_{m=1}^{M-1}\Biggr),\\
\sigma^{(1)}_{f_\Delta}(t \in \mathcal{T}) = \mathrm{WSD}\Biggl(&\Bigl\{\abs{f_m - f_{m+1}} \Bigr\}_{m=1}^{M-1},\\&\Bigl\{\sigma_{f_m} + \sigma_{f_{m+1}} \Bigr\}_{m=1}^{M-1} \Biggr),\\
\end{split}
\end{equation}
where $\mathrm{WSD}(X,\sigma_X)$ denotes the weighted standard deviation of a set of values $X$ with weights $1/\sigma_X^2$, and the superscript ``$(1)$" denotes that this is a measure of the uncertainty due to the fitting quality of each segment alone. We now evaluate the uncertainty due to different potential segmentations of the data.

The segmentation performed by the segmentation algorithm is a function of the auto-segmentation hyperparameters $\lambdall$ and $L_\mathrm{min}$. To estimate the uncertainty due to alternate potential segmentations of the data, we repeat the auto-segmentation with different combinations of hyperparameters chosen from $\lambdall \in \{ 0.9\lambdall^{(o)},0.92\lambdall^{(o)},\ldots,1.08\lambdall^{(o)},1.1\lambdall^{(o)}\}$ and $L_\mathrm{min} \in \{ 0.9L_\mathrm{min}^{(o)},0.92L_\mathrm{min}^{(o)},\ldots,1.08L_\mathrm{min}^{(o)},1.1L_\mathrm{min}^{(o)}\}$, which correspond to the approximate uncertainty in these parameters visible in~\figref{fig:choose_gammma_ll} and~\figref{fig:choose_msl}. The standard deviation for each value of $f_c(t_r)$ and $f_\Delta(t_r)$, $\sigma^{(2)}_{f_c}(t_r)$ and $\sigma^{(2)}_{f_\Delta}(t_r)$ respectively, is the standard deviation of those parameters over all the results of the different hyperparameter combinations.

The total uncertainty is obtained by adding the two contributions:
\begin{equation}
\begin{split}
\sigma_{f_c}(t_r)  &= \sqrt{\sigma^{(1)^2}_{f_c}(t_r) + \sigma^{(2)^2}_{f_c}(t_r) }\\
\sigma_{f_\Delta}(t_r)  &= \sqrt{\sigma^{(1)^2}_{f_\Delta}(t_r) + \sigma^{(2)^2}_{f_\Delta}(t_r) }\\
\end{split}
\end{equation}

\subsubsection{Computing switching rates}
\label{app:subsec:hdfa_switching_rates}
The rates of discrete jumps of fluctuations at a particular timescale are computed from the evaluated $s^{(n)}(t)$ at a particular timescale. Since jumps lasting for durations smaller than the temporal resolution of the method are missed, the rate of observed switching events - denoted as the raw switching rate - does not directly correspond to the true switching rate but can be used to estimate it. The raw rate of switching from state $i$ to state $j$ is computed as 
\begin{equation}
\tilde{\nu}_{i\rightarrow j}^{(n)} = \frac{\sum_k{\delta[s^{(n)}(t_k),i]\delta[s^{(n)}(t_{k+1}),j]}}{\sum_k{(t_{k+1}-t_k)\delta[s^{(n)}(t_k),i]}},
\end{equation}
where $\delta[p,q] = \delta_{p,q}$ is the Kronecker delta function. We now show how we use this to estimate the true switching rate.

To factor these missed fluctuations in the switching rate, it is assumed that discrete jumps that last for time $\tau_{\mathrm{min}}^{(n)} = \delta tL_\mathrm{min}^{(n)}$ or less, where $\delta t$ is the duration of a repetition, are undetected and do not show up in switching of $s^{(n)}(t)$. It is also assumed that the jumps are described by a Poissonian process. Then, the probability of a discrete jump lasting for duration $t>\tau_{\mathrm{min}}^{(n)}$ is $e^{-\tau_{\mathrm{min}}^{(n)} \nu_{i\rightarrow j}^{(n)}}$. The measured raw switching rate is then related to the true switching rate $\nu_{i\rightarrow j}^{(n)}$ by the relation
\begin{equation}
\tilde{\nu}_{i\rightarrow j}^{(n)} = \nu_{i\rightarrow j}^{(n)} e^{-\tau^{(n)}_\mathrm{min} \nu_{i\rightarrow j}^{(n)}}.
\end{equation}
This equation is used to estimate the true switching rate from the measured raw switching rate.

\section{Mapping to physical models}
\label{app:sec:mapping_to_physical_models}

\subsection{Theoretical prediction of \texorpdfstring{$\Delta_\mathrm{CP}$}{dispersion} due to charge parity switching}
\label{app:subsec:theory_cps_max_prediction}
We can obtain the value of $f_{\Delta\mathrm{max}}^{(n)}$ for a fluctuation source corresponding to charge parity switching by using the properties of the qubit. This value is computed both using an analytic approximation of the Transmon Hamiltonian, resulting in $\Delta_\mathrm{CP}^\mathrm{analytical}$, as well as a numerical simulation of the Transmon Hamiltonian, resulting in $\Delta_\mathrm{CP}^\mathrm{numerical}$.

First, we use the qubit frequency and anharmonicity provided in \appref{app:sec:experiment_device_details} to estimate the values of $E_J$ and $E_C$ for each qubit by solving the Transmon Hamiltonian. We obtain $E_C/h = 0.288,0.288,0.287~\mathrm{GHz}$ and  $\xi \equiv E_J/E_C = 43.0, 46.4, 44.1$ for qubits $0, 2, 4$ respectively. We then use the relation~\cite{wilen_correlated_2021,koch_charge-insensitive_2007}
\begin{equation}
    h \Delta_\mathrm{CP}^\mathrm{analytical} = 16 \sqrt{\frac{2}{\pi}} E_C \Biggl(\frac{\xi}{2}\Biggr)^{3/4} e^{-\sqrt{8\xi}} \Biggl[16 \Biggl(\frac{\xi}{2}\Biggr)^{1/2}+1\Biggr]
\end{equation}
to obtain $\Delta_\mathrm{CP}^\mathrm{analytical} = 48.7,26.0,39.7 \kHz$ for the three qubits respectively.

The above relation is valid in the Transmon limit of $E_C/E_J \gg 1$. We can also evaluate the charge dispersion numerically by solving the Transmon Hamiltonian in the plane wave basis. Constructing the Hamiltonian from the estimated $E_C,E_J$ parameters results in the predicted dispersions of $\Delta_\mathrm{CP}^\mathrm{numerical} = 39.8, 21.5, 32.6 \kHz$. 

\subsection{Charge offset shift extraction}
\label{app:subsec:charge_offset_extraction}

In order to evaluate the charge offset $n_g$, we first use the relation
\begin{equation}
    f_{\Delta} (t) = f_{\Delta \mathrm{max}}\abs{\cos(2\pi n_g)}.
\end{equation}

\begin{figure}
    \includegraphics[width=\columnwidth]{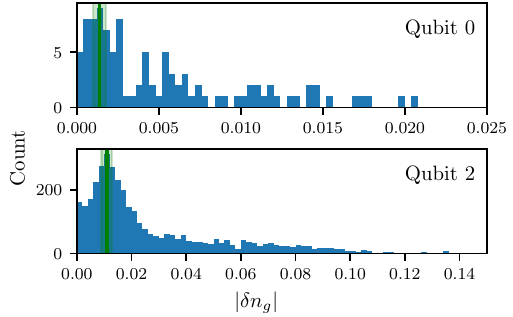}
    \caption{Distribution of the jump in the charge offsets, $|\delta n_g|$, during a jump of the identified dominant TLS for qubits 0 and 2. The location of the highest bar is used as the estimate of the charge offset jump and is depicted by the green vertical lines. The shaded greed region corresponds to the approximate uncertainty in the estimate and is equal to the width of the bars.}
    \label{fig:charge_offset_shifts}
\end{figure}

To evaluate whether the TLS switching leads to a jump in the charge offset, we compare the values of $n_g$ at the time of TLS fluctuations. In \figref{fig:charge_offset_shifts}, we plot the distribution of the absolute value of the difference in charge offset during at the time of a TLS switch, i.e.
\begin{equation}
    |\delta n_g| = |n_g^\mathrm{pre} - n_g^\mathrm{post}|,
\end{equation}
where $n_g^\mathrm{pre}$ and $n_g^\mathrm{post}$ are the values of the charge offset just before and just after a TLS switch.
If the charge offset shift $\delta n_g$ is significantly smaller than the uncertainties in estimates of $n_g$, then a distribution of $\delta n_g$ would have a maximum at $\delta n_g = 0$, with random errors associated with uncertainties in the estimates determining the width of the distribution. However, what is instead observed in~\figref{fig:charge_offset_shifts} is that the distribution has a maximum at $\delta n_g = 0.0014(4)$ and $\delta n_g = 0.011(2)$ for the two qubits respectively, showing that the RTN switching at the second level of the fluctuation hierarchy is correlated with jumps in the qubit charge offset.

\subsection{Derivation of TLS induced charge offset and frequency fluctuations}
\label{app:subsec:charge_dipole_model_derivation}
In this section, we describe the qubit-TLS Hamiltonian and derive the effects on the qubit for the case where the TLS is a defect residing in the barrier which couples to the electric field of the qubit junction~\cite{cole2010quantitative,martinis2005decoherence}. The total Hamiltonian of a qubit interacting with a TLS can be written as
\begin{equation}
    \h = \h_\mathrm{Q} + \h_\mathrm{TLS} + \h_\mathrm{Q-TLS},
\end{equation}
where $\h_\mathrm{Q}$ is the transmon qubit Hamiltonian, $\h_\mathrm{TLS}$ is the TLS Hamiltonian, and $\h_\mathrm{Q-TLS}$ is the qubit-TLS interaction Hamiltonian. The qubit Hamiltonian~\cite{krantz_quantum_2019} is given by 
\begin{equation}
    \h_\mathrm{Q}= 4E_C (\hat{n}-n_g)^2 - E_J \cos(\hat{\phi}),
\end{equation}
where $E_C$ is the charging energy, $\hat{n}$ is the number operator in units of cooper pairs, $n_g$ is the offset charge as before, $E_J$ is the Josephson energy, and $\hat{\phi}$ is the phase operator, with $\left[\hat{\phi}, \hat{n} \right] = i$. The standard tunnelling model (STM) Hamiltonian~\cite{faoro2015interacting,anderson1972anomalous,phillips1987two} for a TLS is given by
\begin{equation}
    \h_\mathrm{TLS} = \frac{1}{2}\epsilon_\mathrm{TLS} \hat{\sigma}_z + \frac{1}{2}\Delta_\mathrm{TLS} \hat{\sigma}_x,
\end{equation}
corresponding to the asymmetry and tunnelling energy respectively, and the TLS operators $\hat{\sigma}_z, \hat{\sigma}_x$ are defined in the basis where the TLS operator for the qubit-TLS interaction is along $\hat{\sigma}_z$. In the considered case of the TLS being a defect residing in the barrier, the charge dipole of the defect interacts with the electric field across the Josephson junction, mediating the qubit-TLS interaction. The interaction Hamiltonian
\begin{equation}
    \h_\mathrm{Q-TLS} = \hat{n}\hat{\sigma}_z\frac{d_\parallel}{2e x}\sqrt{E_C h f_0},
    \label{eq:app_qtls_coupling}
\end{equation}
where $d_\parallel$ is the component of the TLS electric dipole moment parallel to the junction electric field orientation, $x$ is the junction thickness, and $f_\mathrm{0}$ is energy of the qubit in the absence of the TLS~\cite{muller2011coherent,martinis2005decoherence}. 

For clarity, let $\h_\mathrm{Q-TLS} = \lambda \hat{n}\hat{\sigma}_z$. Following~\cite{liu_observation_2024}, we move to the physical basis of the TLS defined by $\hat{\sigma}_x = \cos(\theta)\hat{\eta}_x + \sin(\theta)\hat{\eta}_z, \hat{\sigma}_z = \cos(\theta)\hat{\eta}_z - \sin(\theta)\hat{\eta}_x$, where $\tan(\theta) = \Delta_\mathrm{TLS}/\epsilon_\mathrm{TLS}$. In this basis, the eigenstates of $\hat{\eta}_z$ are the ground and excited states of the TLS, $\ket{g}$ and $\ket{e}$, respectively, and the TLS energy is $h f_\mathrm{TLS} = \sqrt{\epsilon^2_\mathrm{TLS} + \Delta_\mathrm{TLS}^2}$. 
Substituting for $\hat{\sigma}_z$ in \eqref{eq:app_qtls_coupling}, we get

\begin{equation}
\begin{split}
        \h=& 4E_C \Bigl(\hat{n}-n_g + \frac{\lambda\cos(\theta)}{8E_C} \hat{\eta}_z\Bigr)^2 - E_J \cos(\hat{\phi}) \\
        &-\lambda\hat{n}\sin(\theta)\hat{\eta}_x \\
        & - \frac{(\lambda \cos{\theta} \hat{\eta}_z)^2 }{16 E_C} + \lambda n_g \cos(\theta) \hat{\eta}_z\\
        &+ \h_\mathrm{TLS}\\    
        =& 4E_C \Bigl(\hat{n}-n_g + \frac{\lambda\cos(\theta)}{8E_C} \hat{\eta}_z\Bigr)^2 - E_J \cos(\hat{\phi}) \\
        &-\lambda\hat{n}\sin(\theta)\hat{\eta}_x + \h_\mathrm{TLS}^{'},    
\end{split}
\end{equation}
where $ \h_\mathrm{TLS}^{'}$ includes the additional term $\lambda n_g \cos(\theta) \hat{\eta}_z$ with only operators acting on the TLS as well as a constant energy shift $- \frac{(\lambda \cos{\theta})^2}{16 E_C}$ which can be ignored.

Due to the $\lambda \cos(\theta)\hat{\eta}_z$ term, the qubit charge offset has a TLS-state dependent shift. Let $|\delta n_g|$ be the qubit charge offset changes when the TLS state switches from $\ket{g}$ to $\ket{e}$ or vice-versa. After substituting for $\lambda$ and $\cos(\theta)$, we obtain
\begin{equation}
|\delta n_g| = 2\frac{\lambda}{8E_C} \cos(\theta) = \frac{d_\parallel}{8 e x}\sqrt{\frac{h f_0}{E_C}} \frac{\epsilon_{TLS}}{h f_{TLS}}.
\label{eq:app_tls_ng_shift}
\end{equation}

To evaluate the effect of the $\lambda \sin(\theta)$ term, the approximation to the number $\hat{n} \simeq -i(E_J/8E_C)^{1/4}(\hat{a}-\hat{a}^\dagger)/\sqrt{2}$ $=\bra{0}\hat{n}\ket{1}(\hat{a}-\hat{a}^\dagger)$, where $\hat{a}$ and $ \hat{a}^\dagger$ are annihilation and creation operators for a Harmonic oscillator approximation that is valid in the Transmon limit~\cite{koch_charge-insensitive_2007} is considered. With this approximation, the interaction operator $\hat{n}\hat{\eta}_x$ is given by
\begin{equation}
\hat{n}\hat{\eta}_x = \bra{0}\hat{n}\ket{1}(\hat{a}-\hat{a}^\dagger)(\hat{\eta}_+ + \hat{\eta}_-),
\end{equation}
where $\hat{\eta}_\pm$ are the Pauli raising and lowering operators acting on the TLS. Applying the rotating wave approximation leads to dropping of the non-excitation-preserving terms, resulting in
\begin{equation}
\hat{n}\hat{\eta}_x = \bra{0}\hat{n}\ket{1}(\hat{a}\hat{\eta}_+ -\hat{a}^\dagger \hat{\eta}_-).
\end{equation}
Let the corresponding term in the Hamiltonian be $\h_\mathrm{I} = -\lambda\hat{n}\sin(\theta)\hat{\eta}_x $. Then,
\begin{equation}
\begin{split}
\h_\mathrm{I} =&-\lambda \sin(\theta) \bra{0}\hat{n}\ket{1}(\hat{a}\hat{\eta}_+ -\hat{a}^\dagger \hat{\eta}_-)\\
=&g_C(\hat{a}\hat{\eta}_+ -\hat{a}^\dagger \hat{\eta}_-),
\end{split}
\end{equation}
with
\begin{equation}
g_C = -\lambda \sin(\theta) \bra{0}\hat{n}\ket{1}.
\end{equation}
For a non-resonant TLS with $g_C \ll |f_0-f_\mathrm{TLS}|$, this Jaynes-Cummings Hamiltonian leads to a TLS-state dependent shift in the qubit frequency~\cite{krantz_quantum_2019}, $f^{(2)}_\Delta$, of $2 g_C^2/(hf_{12}-f_\mathrm{TLS})$, where $h f_{12} = hf_0 + h\alpha$ is the difference in energy between the 2nd and 1st excited states of the TLS, arising from the coupling induced repulsion between the $\ket{2g}$ and $\ket{1e}$ states~\cite{liu_observation_2024}. This results in
\begin{equation}
\begin{split}
f^{(2)}_\Delta =& \frac{2g_C^2}{h (f_0 + \alpha - f_{TLS})}\\
=& 2 \frac{\lambda^2 \sin(\theta)^2 \bra{0}\hat{n}\ket{1}^2}{h (f_0  + \alpha- f_{TLS})} \\
=& 2 \Biggl[\frac{d_\parallel}{2e x}\sqrt{E_C h f_0}\Biggr]^2 \Bigl( \frac{\Delta_\mathrm{TLS}}{h f_\mathrm{TLS}}\Bigr)^2\frac{ \bra{0}\hat{n}\ket{1}^2}{h (f_0 + \alpha - f_\mathrm{TLS})} \\
=& \Biggl[\frac{d_\parallel}{e x}\bra{0}\hat{n}\ket{1} \frac{\Delta_\mathrm{TLS}}{h f_\mathrm{TLS}}\Biggr]^2\frac{E_C f_0}{2 (f_0 + \alpha - f_\mathrm{TLS})}. \\
\label{eq:app_tls_dispersive_shift}
\end{split}
\end{equation}

The TLS is represented by two potential wells and tunnelling between the two potential wells. The two effects of the TLS switching on the qubit, namely the charge offset jump and the frequency shift, are affected by the asymmetry and tunnelling energy of the TLS. The TLS is Since $|\delta n_g|$ depends on the change in the charge configuration, the more localised the TLS wavefunctions are on each potential well, the greater the effect on $n_g$. This explains the origin of $\frac{\epsilon_\mathrm{TLS}}{h f_\mathrm{TLS}}$ in the relation for $|\delta n_g|$ since a larger $\epsilon_\mathrm{TLS}$ relative to $hf_\mathrm{TLS}$ corresponds to more localised TLS wavefunctions. On the other hand, the centre frequency shift relies on hybridization between the qubit and TLS energy levels arising from the quantum mechanical coupling between the two systems rather than just switching between two charge distributions. Therefore, the centre frequency shift depends on the tunnelling component of the TLS $\frac{\Delta_\mathrm{TLS}}{h f_\mathrm{TLS}}$.

In \eqref{eq:app_tls_ng_shift} and \eqref{eq:app_tls_dispersive_shift}, the independent parameters involved in the two equations are $|\delta n_g|$, $d_\parallel$, $x$, $f_0$, $\alpha$, $\epsilon_\mathrm{TLS}$, $E_C$, $f_\Delta^{(2)}$, $\bra{0}\hat{n}\ket{1}$, $f_\mathrm{TLS}$. The parameters $|\delta n_g|$ and $f_\Delta^{(2)}$ are obtained from the HDFA results. The parameters $f_0$ and $\alpha$ are directly obtained from the device calibration data and $E_C$ is evaluated from $f_0$ and $\alpha$ as before. The matrix element $\bra{0}\hat{n}\ket{1}$ is obtained from numerical simulation of the Transmon Hamiltonian $\h_Q$. The remaining unknown parameters are $d_\parallel,x,\epsilon_\mathrm{TLS}$,$f_\mathrm{TLS}$. Since there are two equations, if the junction thickness and TLS energy $h f_\mathrm{TLS}$ are known then the system of equations can be solved and the remaining parameters identified.

In the absence of measurements of the junction thickness, we assume the thickness to be between $1~\mathrm{nm}<x < 2~\mathrm{nm}$ as observed in prior literature~\cite{Zeng_2015}. The remaining unknown parameter is the TLS energy. Avoided crossings in a qubit spectroscopy experiment or dips in qubit coherence upon frequency tuning can be used to evaluate the energies of the TLS~\cite{lisenfeld_electric_2019,Klimov_2018}. In the absence of capabilities or access to perform such experiments, the TLS energy can be estimated from the observed fluctuation rates of the TLS assuming that it is in thermal equilibrium such that the population of its ground and excited states are given by the Boltzmann distribution. Using the balance equation
\begin{equation}
\frac{\nu^{(2)}_{0\rightarrow 1}}{\nu^{(2)}_{1\rightarrow 0}} = e^{-\frac{h f_\mathrm{TLS}}{k_\mathrm{B}T_\mathrm{TLS}}},
\end{equation}
where $T_\mathrm{TLS}$ is the effective temperature of the TLS environment, the TLS energy is given by 
\begin{equation}
hf_\mathrm{TLS} = k_\mathrm{B}T_\mathrm{TLS} \ln {\frac{\nu^{(2)}_{1\rightarrow 0}}{\nu^{(2)}_{0\rightarrow 1}}}.
\end{equation}
Although $T_\mathrm{TLS}$ is not known, it expected to be greater the dilution refrigerator temperature of about $10~\mathrm{mK}$ and assumed to be below $100~\mathrm{mK}$ based on previous measurements in literature~\cite{lucas2023}. For this range of temperatures, the obtained TLS energies are $f_\mathrm{TLS} = 0.38-3.8~\mathrm{GHz}$ and $0.16- 1.6~\mathrm{GHz}$ for the TLS coupled to qubits 0 and 2, respectively. 

With these ranges for the junction thickness $x$ and the TLS energy $f_\mathrm{TLS}$, \eqref{eq:app_tls_ng_shift} and \eqref{eq:app_tls_dispersive_shift} can be used to compute $d_\parallel, \epsilon_\mathrm{TLS}$ and $\Delta_\mathrm{TLS}$. 